\documentclass[preprint]{aastex}
\usepackage{lscape}

\begin{document}
  \title{Chasing the Chelyabinsk asteroid \textit{N}-body style}

  \author{C.~de~la~Fuente~Marcos and R. de la Fuente Marcos}
   \affil{Apartado de Correos 3413, E-28080 Madrid, Spain}
   \email{carlosdlfmarcos@gmail.com}

  \and

  \author{S. J. Aarseth}
   \affil{Institute of Astronomy, University of Cambridge,
         Madingley Road, Cambridge CB3 0HA, UK}

  \begin{abstract}
     On 2013 February 15 a small asteroid rammed against the atmosphere above the region 
     of Chelyabinsk in Russia, producing the most powerful superbolide since the Tunguska 
     event in 1908. Lacking proper astrometric observations, the pre-impact orbit of this 
     object has been determined using videos, satellite images, and pure geometry. 
     Unfortunately, more than two years after the event, the published estimates vary so 
     much that there is no clear orbital solution that could be used to investigate the 
     origin of the impactor and the existence of dynamically, or perhaps even genetically, 
     related asteroids. Here, we revisit this topic using a full $N$-body approach. A 
     robust statistical test is applied to published solutions to discard those unable to 
     produce a virtual impact at the observed time (03:20:20.8$\pm$0.1~s~UTC). The same 
     $N$-body methodology and the latest ephemerides are used to compute a new orbital 
     solution: $a$~=~1.6247~AU, $e$~=~0.5318, $i$~=~3$\fdg$9750, $\Omega$~=~326$\fdg$4607 
     and $\omega$~=~109$\fdg$7012. This new solution ---which has an impact probability 
     $>$~0.99999 and uncertainties in time and space of 0.2~s and 6~km, respectively--- is
     utilized to explore the past orbital evolution of the impactor as well as the 
     presence of near-Earth objects moving in similar paths. A dynamical link between 
     asteroid 2011~EO$_{40}$ and the Chelyabinsk impactor is confirmed. Alternative 
     orbital solutions are extensively explored.
  \end{abstract}


  \section{Introduction}
     Asteroids diving out of the Sun's blinding glare represent a very real threat that cannot be easily detected or defended against with 
     currently available resources. In general, any minor body that encounters our planet after reaching perihelion inside the orbit of the 
     Earth will approach undetected, hidden in the daytime sky. If the true minimal approach distance to our planet is small enough (e.g., 
     $<$0.001 AU) this configuration is sometimes described as the Red Baron dynamical scenario (Adamo 2011), in which the glare of the Sun 
     provides an effective hiding spot to a close passage or an eventual impactor. The most dramatic recent example of such an occurrence 
     was the Chelyabinsk event. 

     On 2013 February 15 a small asteroid approached the Earth undetected, coming from the direction of the Sun, and slammed against the 
     atmosphere above the region of Chelyabinsk in Russia, producing the most powerful superbolide since the Tunguska event in 1908 (Brown 
     et al. 2013; Le Pichon et al. 2013). In the absence of pre-impact astrometric observations, the orbit followed by the parent body of 
     the Chelyabinsk superbolide has been determined using videos recorded by witnesses (Borovi\v{c}ka et al. 2013; Popova et al. 2013; 
     Zuluaga et al. 2013; Emel'Yanenko et al. 2014; Dmitriev et al. 2015; Golubaev 2015), satellite images (Proud 2013), and pure geometry 
     (de la Fuente Marcos \& de la Fuente Marcos 2013, 2014; hereinafter Papers I and II, respectively). Unfortunately, more than two years 
     after the event, the published estimates vary so much that there is no clear orbital solution that could be used to investigate the 
     origin of the impactor and the existence of dynamically and/or physically related asteroids.

     The notion of pre-impact orbit is widely used in the study of cosmic strikes, but, what is the pre-impact orbit of the parent body of a 
     meteor? The most natural answer to this non-trivial question would be that it is the one computed for an epoch sufficiently distant 
     from the impact time and such that its subsequent evolution produces an impact within a reasonable time frame imposed by the available 
     observational data. Such an accurate orbital solution is appropriate to study both the past dynamical evolution of the parent body of 
     the meteor and the possible existence of other small bodies moving in similar orbits among known near-Earth objects (NEOs). No matter 
     the methodology behind a given orbital solution, it must comply with the hard experimental evidence: impact time and location. Orbital 
     solutions must be rejected on the grounds of their failure to generate an impact at the correct time on the right spot. In the case of 
     the Chelyabinsk superbolide, it was first detected on 2013 February 15 03:20:20.8$\pm$0.1 s UTC ($t_{\rm impact}$) at longitude 
     64\fdg565$\pm$0\fdg030 ($\lambda_{\rm impact}$), latitude +54\fdg445$\pm$0\fdg018 ($\phi_{\rm impact}$) and altitude 97.1$\pm$0.7 km 
     (see Table S1, Popova et al. 2013, also Borovi\v{c}ka et al. 2013; Miller et al. 2013). 

     Here, we revisit the topic of the pre-impact orbit of the Chelyabinsk superbolide using a full $N$-body methodology to both rank the 
     published solutions and obtain a new one. This paper is organized as follows. The published orbits are briefly discussed in Section 2. 
     In Section 3, a statistically robust impact test is described, validated, and applied to published orbital solutions. Our full $N$-body 
     approach is presented in Section 4. A new pre-impact orbit is derived in Section 5; alternative orbital solutions are also extensively 
     explored there. This new orbital solution is used in Section 6 to study the past dynamical evolution of the Chelyabinsk impactor and 
     investigate the presence of known NEOs moving in similar orbits. Results are discussed and conclusions summarized in Section 7.

  \section{The pre-impact orbit so far}
     More or less detailed orbital solutions have been presented in Adamo (2013), Borovi\v{c}ka et al. (2013), Chodas \& 
     Chesley,\footnote{http://neo.jpl.nasa.gov/news/fireball\_130301.html} Dmitriev et al. (2015), Emel'Yanenco et 
     al.,\footnote{http://www.inasan.ru/eng/asteroid\_hazard/chelyabinsk\_bolid\_new.html} Emel'Yanenko et al. (2014), Golubaev (2015), 
     Green (2013), Lyytinen,\footnote{http://www.amsmeteors.org/2013/02/large-daytime-fireball-hits-russia/} Lyytinen et 
     al.,\footnote{http://www.projectpluto.com/temp/chelyab.htm} Nakano,\footnote{http://www.icq.eps.harvard.edu/CHELYABINSK.HTML} Popova et 
     al. (2013), Proud (2013), Zuluaga \& Ferrin,\footnote{http://arxiv.org/abs/1302.5377} Zuluaga et al. 
     (2013),\footnote{http://astronomia.udea.edu.co/chelyabinsk-meteoroid/} and Papers I and II. The vast majority of these solutions have 
     been obtained from videos recorded by witnesses, but satellite images (Proud 2013) and pure geometry (Papers I and II) have also been 
     used. All of them show that the dynamical class of the Chelyabinsk impactor is Apollo and the impact occurred at the descending node,
     but the actual values of the orbital elements are still under dispute (see Table \ref{orbits} for values of semimajor axis, $a$, 
     eccentricity, $e$, inclination, $i$, longitude of the ascending node, $\Omega$, argument of perihelion, $\omega$, and time of 
     perihelion passage, $\tau$; the rest can be found in Papers I and II). The relative dispersion between solutions is large enough to 
     make them incompatible: 10.9\% in $a$, 9.3\% in $e$, 25.4\% in $i$, 0.025\% in $\Omega$, and 6.4\% in $\omega$. In the following, we 
     apply a new technique ---using full $N$-body calculations--- to test published orbital determinations statistically and derive a robust 
     solution.
%
%
     \begin{landscape}
     \begin{table}
      \centering
      \fontsize{8}{11pt}\selectfont
      \tabcolsep 0.05truecm
      \caption{Published Solutions for the Pre-impact Orbit of the Chelyabinsk Impactor}
      \begin{tabular}{lccccccc}
       \hline
          Authors                     & $a$ (AU)          & $e$               & $i$ (deg)        & $\Omega$ (deg)        & $\omega$ (deg)        & 
          $\tau$ (JDCT)          & $P_{\rm imp}$  \\
       \hline
          Green (2013)                & 1.55$\pm$0.07     & 0.50$\pm$0.02     & 3.6$\pm$0.7      & 326.410$\pm$0.005     & 109.7$\pm$1.8         &
          --                     & --             \\
          Borovi\v{c}ka et al. (2013) & 1.72$\pm$0.02     & 0.571$\pm$0.006   & 4.98$\pm$0.12    & 326.459$\pm$0.001     & 107.67$\pm$0.17       &
          2456292.89$\pm$0.17    & $\leq 10^{-5}$ \\
          Dmitriev et al. (2015)      & 1.76$\pm$0.04     & 0.580$\pm$0.012   & 5.0$\pm$0.3      & 326.454$\pm$0.002     & 108.2$\pm$0.7         &
          --                     & --             \\
          Emel'Yanenko et al. (2014)  & 1.88$\pm$0.07     & 0.609$\pm$0.017   & 5.9$\pm$0.4      & 326.446$\pm$0.002     & 108.9$\pm$0.5         &
          2456294.4$\pm$5.4      & $<    10^{-6}$ \\
          Golubaev (2015)             & 1.67$\pm$0.10     & 0.57$\pm$0.03     & 7.1$\pm$0.5      & 326.42                & 106.3$\pm$2.5         &
          --                     & --             \\
          Paper I                     & 1.62375$\pm$0.00014 & 0.53279$\pm$0.00011 & 3.817$\pm$0.005  & 326.4090$\pm$0.0007   & 109.44$\pm$0.03   &
          2456292.478$\pm$0.005  & $\leq$ 0.005   \\
          Paper II                    & 1.624765$\pm$0.000005 & 0.53184$\pm$0.00001 & 3.97421$\pm$0.00005  & 326.44535$\pm$0.00001   & 109.71442$\pm$0.00004       &
          2456292.57800$\pm$0.00004 & $\leq$ 0.77 \\
          Popova et al. (2013)        & 1.76$\pm$0.08     & 0.581$\pm$0.009   & 4.93$\pm$0.24    & 326.4422$\pm$0.0014   & 108.3$\pm$1.9         &
          2456293.4$\pm$2.0      & $\leq 10^{-6}$ \\
          Proud (2013)                & 1.47$^{+0.03}_{-0.13}$ & 0.52$^{+0.01}_{-0.05}$ & 4.61$^{+2.58}_{-2.09}$ & 326.53$^{+0.01}_{-0.0}$ & 96.58$^{+2.94}_{-1.72}$ &
          --                     & --             \\
          Zuluaga \& Ferrin$^6$       & 1.73$\pm$0.23     & 0.51$\pm$0.08     & 3.45$\pm$2.02    & 326.70$\pm$0.79       & 120.62$\pm$2.77       &
          --                     & --             \\
          Zuluaga et al. (2013)       & 1.27$\pm$0.05     & 0.44$\pm$0.02     & 3.0$\pm$0.2      & 326.54$\pm$0.08       & 95.1$\pm$0.8          &
          --                     & --             \\
          Zuluaga et al.$^7$          & 1.368$\pm$0.006   & 0.470$\pm$0.010   & 4.0$\pm$0.3      & 326.479$\pm$0.003     & 99.6$\pm$1.3          &
          --                     & --             \\
       \hline
          \,\,\,\, Average$\pm\sigma$ & 1.6$\pm$0.2       & 0.53$\pm$0.05     & 4.5$\pm$1.1      & 326.48$\pm$0.08       & 107$\pm$7             &
          2456293.1$\pm$0.4      &                \\
       \hline
          This work                   & 1.62470348              & 0.53184268              & 3.9749908              & 326.4607324             & 109.7012184                   &
           2456292.5834112           &          \\
                                      &          $\pm$          &          $\pm$          &         $\pm$          &          $\pm$          &           $\pm$               &
                        $\pm$        & $>$0.99999 \\
                                      &              0.00000005 &              0.00000002 &              0.0000008 &               0.0000014 &                0.0000008      &
                           0.0000010 &          \\
       \hline
      \end{tabular}
      \label{orbits}
      \tablenotetext{}{Note. The errors from Popova et al. (2013) are from their Table S5B. $P_{\rm imp}$ is the impact probability. Paper I 
                       is de la Fuente Marcos \& de la Fuente Marcos (2013). Paper II is de la Fuente Marcos \& de la Fuente Marcos (2014).}
     \end{table}
     \end{landscape}
%
%

  \section{Orbit validation: impact test}
     Independently from the methodology used to derive it, any computed orbit must be consistent with the well-established fact that a 
     superbolide was observed on 2013 February 15 03:20:20.8$\pm$0.1 s UTC at longitude 64\fdg565$\pm$0\fdg030, latitude 
     +54\fdg445$\pm$0\fdg018, and altitude 97.1$\pm$0.7 km (see Table S1, Popova et al. 2013). Our solution in Paper II satisfies these 
     constraints relatively well, but it was referred to an epoch arbitrarily close to that of the impact event; therefore, it does not 
     describe the pre-impact orbit of the Chelyabinsk impactor in a strict sense. The same can be said about other published solutions. 

     \subsection{Method}
        The robust statistical impact test described in this section can be viewed as an independent implementation of the ideas explored in 
        Sitarski (1998, 1999, 2006).\footnote{http://phas.cbk.waw.pl/neo.htm} In his work, Sitarski uses pre-impact observational 
        information as input to develop his methodology to predict asteroid impacts. Any asteroid's full orbital solution can be used to 
        generate synthetic observational data suitable to be tested for virtual impacts or close encounters with our planet or any other
        body in the solar system. The approach is very simple; we assume a set of orbital elements ($a$, $e$, $i$, $\Omega$, $\omega$, and 
        $\tau$) at a given epoch $t_0$, generate Cartesian state vectors ($\textit{\textbf{r}}$ and $\textit{\textbf{v}}$) for the assumed 
        orbit at the reference epoch, and use $N$-body simulations within a certain physical model to study the evolution of the assumed 
        orbit until an impact or a miss occurs. If a large sample of orbits is studied, the conventional statistical analysis of their 
        outcomes in the form of frequency histograms (for $t_{\rm impact}$ and other parameters) should be enough to decide if a candidate 
        impact orbit is statistically robust or not. In our case, a Monte Carlo approach (Metropolis \& Ulam 1949; Press et al. 2007) is 
        used to generate sets of orbital elements. Unless explicitly stated, Gaussian random numbers are utilized to emulate better the 
        results from traditional, astrometry-based orbital solutions; the Box-Muller method is applied to generate random numbers with a 
        normal distribution (Press et al. 2007).

        Our model solar system includes the perturbations from the eight major planets and treats the Earth and the Moon as two separate 
        bodies. It also incorporates the barycenter of the dwarf planet Pluto--Charon system and the five most massive asteroids of the main 
        belt, namely, (1) Ceres, (2) Pallas, (4) Vesta, (10) Hygiea, and (31) Euphrosyne. Using a different number of perturbing asteroids 
        has no impact on most of our results (impact tests and pre-impact orbit determinations). However, a small but measurable variation  
        is found when investigating the past evolution of the various objects studied here for the time interval considered; the variations 
        are too small to affect any of our conclusions. We use initial conditions (positions and velocities referred to the barycenter of 
        the solar system) provided by the Jet Propulsion Laboratory (JPL) \textsc{horizons} system (Giorgini et al. 1996; Standish 1998) and 
        relative to the Julian Date 2456337.638888889 (= A.D. 2013 February 14 03:20:00.0000), Coordinate Time (JDCT) epoch ($t_{0}$, $t$ = 
        0 in the figures, see Table \ref{Cartesian}), i.e., the integrations are started $\sim$24 hr before $t_{\rm impact}$. We retain the 
        level of precision in time ($\sim$0.0001 s) provided by the \textsc{horizons} system throughout the paper. Cartesian state vectors 
        for the test orbits are generated using the Monte Carlo technique pointed out above, within some given or assumed ranges for the 
        orbital parameters, and the usual expressions in, e.g., Murray \& Dermott (1999). The $N$-body simulations performed here were 
        completed applying the Hermite integration scheme described by Makino (1991) and implemented by Aarseth (2003). The standard version 
        of this direct $N$-body code is publicly available from the IoA web site.\footnote{http://www.ast.cam.ac.uk/$\sim$sverre/web/pages/nbody.htm} 
        Non-gravitational forces, relativistic and oblateness terms are not included in the calculations; additional details can be found in 
        de la Fuente Marcos \& de la Fuente Marcos (2012), where the results of this $N$-body code are compared with those from other codes 
        as well. For the case studied here, the role of the Earth's oblateness is rather negligible ---see the analysis in Dmitriev et al. 
        (2015). Relative errors in the total energy are as low as 10$^{-14}$ to 10$^{-13}$ or lower. The relative error in the total angular 
        momentum is several orders of magnitude smaller. 

        In the particular problem that we are considering here, our choice of $t_{0}$ places the impactor near the edge of the Hill sphere 
        of the Earth (0.0098 AU) at the beginning of the simulation (0.008 AU) and results in a systematic difference at the end of the 
        integration, between our ephemerides and those provided by the JPL for the Earth, of about 6 km. As the average orbital speed of our 
        planet is 29.78 km s$^{-1}$, it implies that the temporal systematic error in our impact calculations could be as small as 0.2 s, 
        which matches well the actual uncertainty in $t_{\rm impact}$. In comparison, the time taken by our planet to travel a distance 
        equal to its own average diameter (12,742 km, $R_{\rm E}$ = 6371 km) is nearly 7.1 minutes. A spatial error of 6 km is equivalent 
        to an angular error of 0\fdg054 in geographical coordinates that also parallels the level of angular precision in $\lambda_{\rm 
        impact}$ and $\phi_{\rm impact}$. Therefore, our results are as realistic as they can possibly be within the known observational 
        uncertainties. 

        The very precise values of the impact parameters available for this particular impact event impose very tight limits on the maximum 
        values of the systematic errors that can be tolerated during the integrations in order to obtain statistically meaningful results. 
        For example, if the value of the integration errors in the position of the Earth at the end of the simulations is of the order of a 
        few hundred kilometers, this is equivalent to an error in the angular quantities $>1$\degr and therefore more than 30 times the 
        largest deviation in impact coordinates; such a scenario is completely unacceptable under the present circumstances and cannot lead 
        to any usable results. This important issue has been neglected in the published literature on this subject although it is of the 
        utmost importance in this particular case. 

     \subsection{Quality control: the case of Duende}
        Before applying the $N$-body-based statistical test to any solution, one key question must be answered: How reliable is the test? 
        Can we trust its results? Perhaps it includes some kind of unknown systematics that may favor some solutions over others or, as 
        pointed out above, the computational errors are large enough to produce inconclusive results. Synthetic data, generated under 
        controlled conditions, are often used to validate statistical tests before applying them to real data. In our case, this approach 
        may add more sources of uncertainty as the test itself rests on generating a very large amount of synthetic data (but nonetheless 
        based on observational data). Being able to test our approach with data coming from a well-studied real event would be far more 
        advantageous. 

        Fortunately or not, the Chelyabinsk event was not the only spectacular cosmic event that took place on 2013 February 15: asteroid 
        367943 Duende (2012 DA$_{14}$) passed nearly 27,700 km above the Earth's surface, well inside the boundaries of the ring of 
        geosynchronous satellites although almost perpendicular to it, reducing the chances of an actual collision with one of them. This 
        rather unusual episode had been expected for about a year (see, e.g., Wlodarczyk 2012) and it was followed closely by the 
        scientific community worldwide (see, e.g., de Le\'on et al. 2013; Nechaeva et al. 2013; Terai et al. 2013; Urakawa et al. 2013; 
        Takahashi et al. 2014). While waiting for this close encounter to happen, the Chelyabinsk event took place. The evidence compiled 
        so far indicates that the two events were completely independent and unrelated. 

        Duende was closest to the Earth on February 15 at approximately 19:25:49.4 UTC ($t_{\rm close}$ = 2456339.309600038$\pm$0.000000120 
        JDCT), 27,679$\pm$15 km above the Earth's surface.\footnote{http://www.nasa.gov/topics/solarsystem/features/asteroidflyby.html}$^{,}$ 
        \footnote{http://ssd.jpl.nasa.gov/sbdb.cgi?sstr=2012DA14;cad=1\#cad} At the time of closest approach, the asteroid flew over the 
        eastern Indian Ocean, $\lambda\sim97\fdg5$ E and $\phi\sim6$\degr S, off the Indonesian island of Sumatra. In order to try to 
        reproduce these close encounter parameters (and their errors) derived by the JPL's Solar System Dynamics Group (SSDG), we use the 
        tools and numerical model described above and analyze the results. In principle, the orbital elements of the test orbits can be 
        obtained by varying them randomly, within the ranges defined by their mean values and standard deviations (i.e., as provided by the 
        \textsc{horizons} system). For example, a new value of the orbital inclination can be found using the expression 
        $i_{\rm t} = \langle{i}\rangle + \sigma_{i}\,r_{\rm i}$, where $i_{\rm t}$ is the inclination of the test orbit, $\langle{i}\rangle$ 
        is the mean value of the inclination (candidate pre-impact orbit or full orbit from the \textsc{horizons} system), $\sigma_{i}$ is 
        the standard deviation of $i$, and $r_{\rm i}$ is a (pseudo) random number with normal distribution in the range $-$1 to 1. Sitarski 
        (1998, 1999, 2006) has pointed out that this is equivalent to considering a number of different virtual minor planets moving in 
        similar orbits, but not a sample of test orbits incarnated from a set of observations obtained for a single minor planet. The orbit 
        of Duende has been computed from a set of observations and, therefore, applying the classical ---but statistically wrong--- 
        expressions may lead to unphysical results. The correct statistical alternative is to consider how the elements affect each other, 
        applying the Monte Carlo using the Covariance Matrix (MCCM) approach (Bordovitsyna et al. 2001; Avdyushev \& Banschikova 2007), or 
        to follow the procedure described in Sitarski (1998, 1999, 2006). 

\begin{figure}
        \centering
        \includegraphics[width=\linewidth]{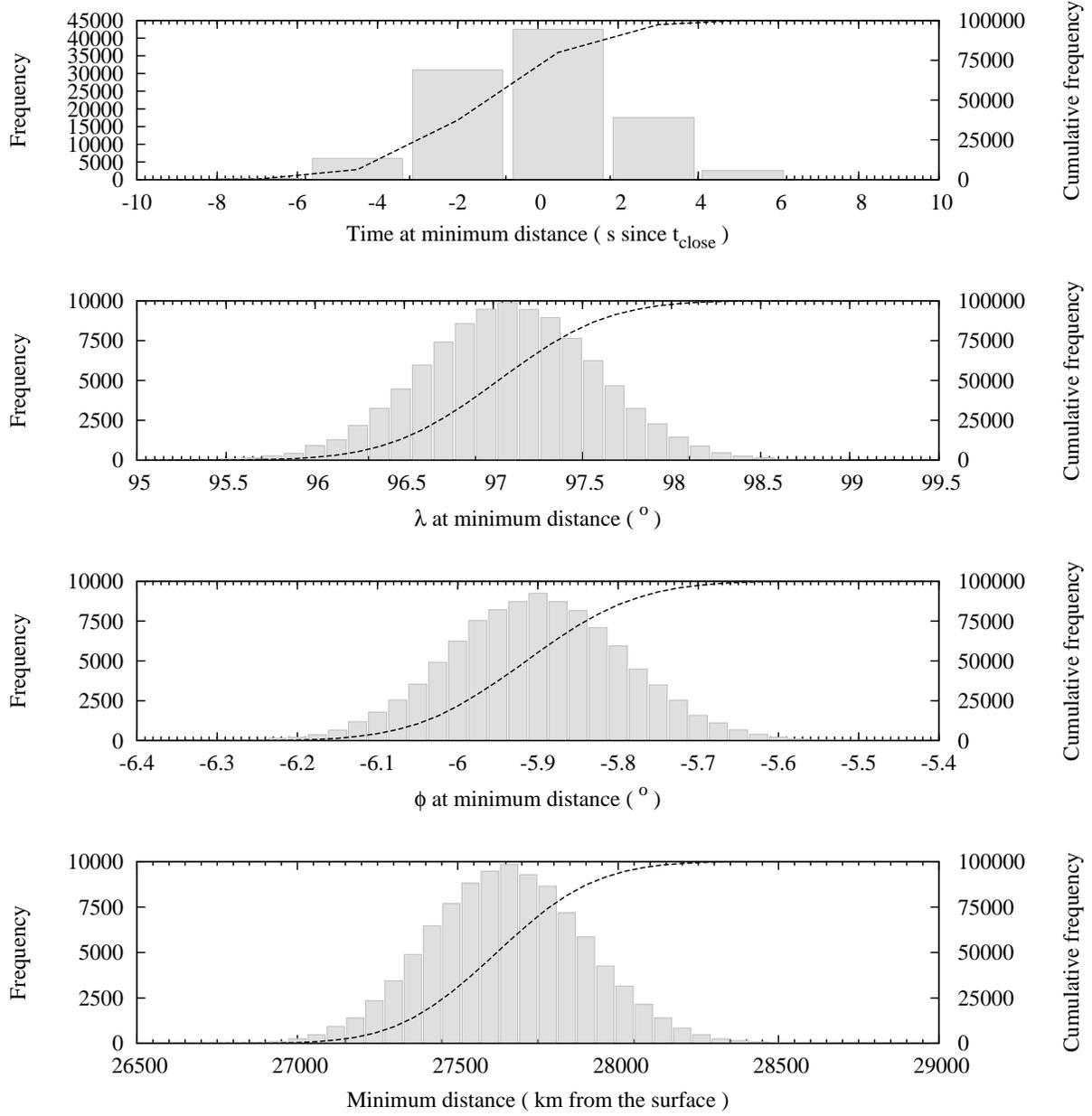}
        \caption{Parameters of the close encounter with 367943 Duende (2012 DA$_{14}$) on 2013 February 15 for a set of 10$^{5}$ numerical
                 experiments using initial conditions generated by applying the classical approach described in the text. The value of 
                 $t_{\rm close}$ as determined by the JPL's SSDG is 2456339.309600038$\pm$0.000000120 JDCT at a minimum distance of 
                 27,679$\pm$15 km above the Earth's surface. The number of bins in the top panel is small because of the unavoidable 
                 discretization of the output interval.
                }
        \label{duende}
     \end{figure}

\begin{figure}
        \centering
        \includegraphics[width=\linewidth]{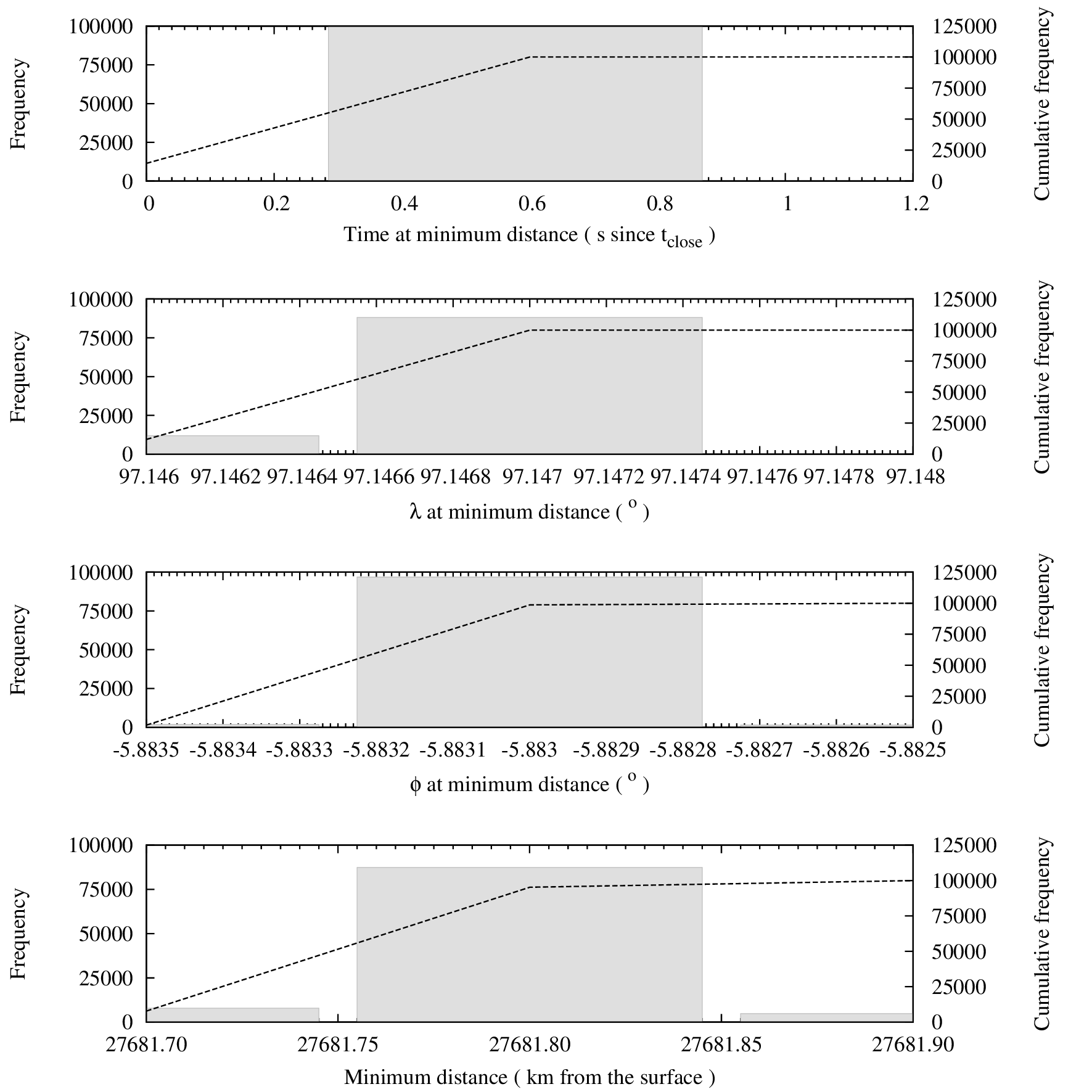}
        \caption{Same as Figure \ref{duende} but for numerical experiments using initial conditions generated by applying the covariance 
                 matrix (see the text for details). The bins are few because the values of the dispersions are very small and due to the 
                 unavoidable discretization of the output interval. The difference between our value for the time of closest approach and
                 the one determined by the JPL's SSDG is 0.627 s and that of the distance of closest approach is 2.9 km. These values are 
                 consistent with the level of precision required to conduct the statistical test discussed in the text.}
        \label{duendex}
     \end{figure}

        Figure \ref{duende} displays the results of a set of 10$^{5}$ numerical experiments using initial conditions generated by applying 
        the classical but wrong approach that neglects the covariance matrix. It is clear that the integrations reproduce the data obtained 
        by the SSDG but the dispersions are rather large. To further explore this remarkable close encounter, we have used an implementation 
        of the MCCM approach (for full mathematical details see Section 3 in de la Fuente Marcos \& de la Fuente Marcos 2015); i.e., a Monte 
        Carlo process creates test orbits with initial parameters from the nominal orbit (for the $t_{0}$ epoch), adding random noise on 
        each initial orbital element and making use of the covariance matrix. The results for this new set of 10$^{5}$ numerical experiments 
        are presented in Figure \ref{duendex}. Consistently with Figure 5 in Sitarski (1998), the outcomes from these two approaches are 
        very different but nonetheless statistically compatible. The difference between our value for the time of closest approach 
        (2456339.309607290 JDCT) using MCCM and the one determined by the SSDG is 0.627 s, and that of the distance of closest approach 
        (27,681.80 km) is 2.9 km. Regarding the relative velocity at closest approach, the value quoted by the SSDG (no errors) is 
        7.81996942783692~km~s$^{-1}$ and the one from our MCCM approach is 7.8199012$\pm$0.0000007~km~s$^{-1}$. In this and future 
        calculations the error quoted is the standard deviation (1$\sigma$) unless explicitly stated. These results match the level of 
        precision required to conduct the statistical test discussed in the previous section. The methodology described above is robust 
        enough to provide objective and reliable results. If any of the tested solutions is unable to generate virtual impacts consistent 
        with the observational data, this will not be due to the test itself, but because the actual pre-impact candidate orbital solution 
        is incorrect. In this context, any solution giving a reasonable fraction of impacts with parameters within the observational 
        uncertainties can be considered as a robust pre-impact orbital solution.
%
%
%
%
%
%
%
%

     \subsection{Impact test: results}
        In order to assess the suitability of a given solution by applying the methodology described in the previous sections, we require 
        the entire set of six orbital elements and their standard deviations, which are only readily available for the orbital solutions 
        presented in Borovi\v{c}ka et al. (2013) and Popova et al. (2013). These two solutions are regarded by many as the best 
        determinations published so far (but see Paper II, which presents a different statistical test that is applied to most of the 
        orbital solutions in Table \ref{orbits}). For these orbital solutions and the one in Emel'Yanenko et al. (2014) ---which is 
        presented as an improvement with respect to Popova et al. (2013)--- we have studied the evolution of $10^{6}$ test orbits from 
        $t_{0}$ until some time after $t_{\rm impact}$ using the scheme outlined above. The orbital elements of the test orbits for a given 
        orbital solution have been computed, varying them randomly, within the ranges defined by their mean values and standard deviations 
        (see Table \ref{orbits}) as described above. No covariance matrices have been used in these tests because none has been included 
        among the results published so far. The only differences between the computations described here and those carried out for Duende in 
        the previous section are in the input average values and their standard deviations (see Table \ref{orbits}) for the various orbital 
        elements, and in the fact that the simulated time in the case of Duende is nearly 16 hr longer.

\begin{figure}
        \centering
        \includegraphics[width=0.5\linewidth]{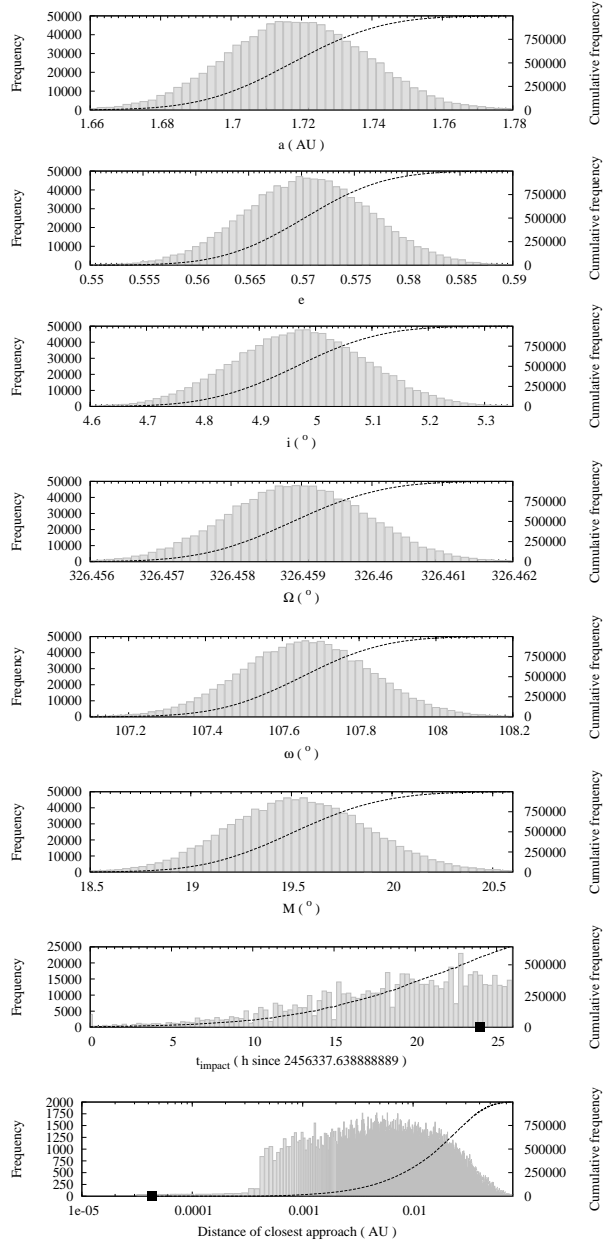}
        \caption{Distribution in orbital parameter space for an experiment using the orbital solution presented in Borovi\v{c}ka et al. 
                 (2013); the impact time ($\sim$24 hr after $t$ = 0) according to Popova et al. (2013) and the upper atmosphere limit (115 
                 km) are indicated as black squares. This figure is the result of the evolution of $10^6$ test orbits. The resulting 
                 distributions in impact/close-encounter parameter space are displayed in the bottom and second-to-bottom panels. Nearly 
                 30\% of the test orbits reach their minimal distance to our planet at the end of the integration; this is why that 
                 fraction is missing from the cumulative distribution in the impact time panel. The first six panels provide the input 
                 distributions.}
        \label{borovicka}
     \end{figure}

\begin{figure}
        \centering
        \includegraphics[width=0.6\linewidth]{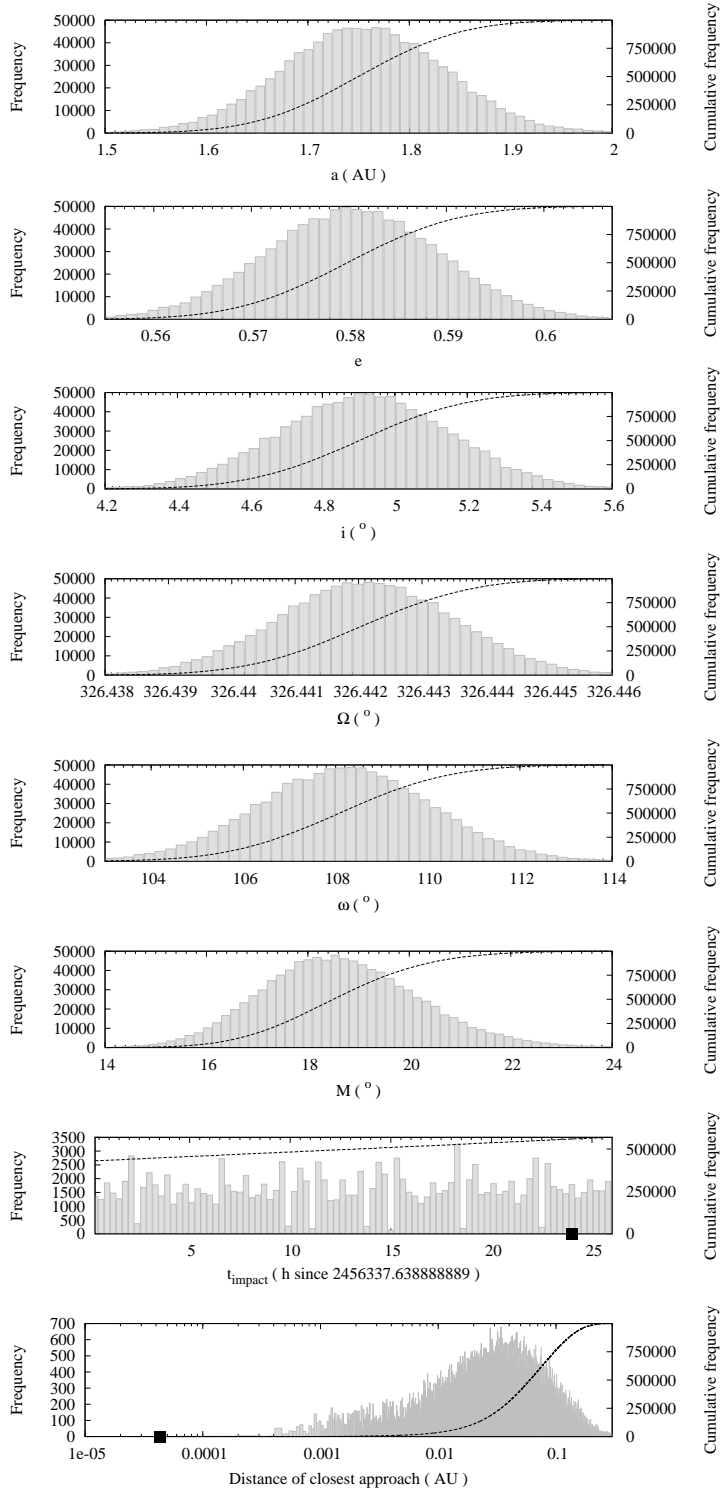}
        \caption{Outcome of an experiment similar to that in Figure \ref{borovicka} but for the orbital solution described in Popova et al. 
                 (2013). Only about 10\% of the orbits tested reach their minimal distance to the Earth during the time interval displayed 
                 in the impact time panel.}
        \label{popova}
     \end{figure}

\begin{figure}
        \centering
        \includegraphics[width=0.6\linewidth]{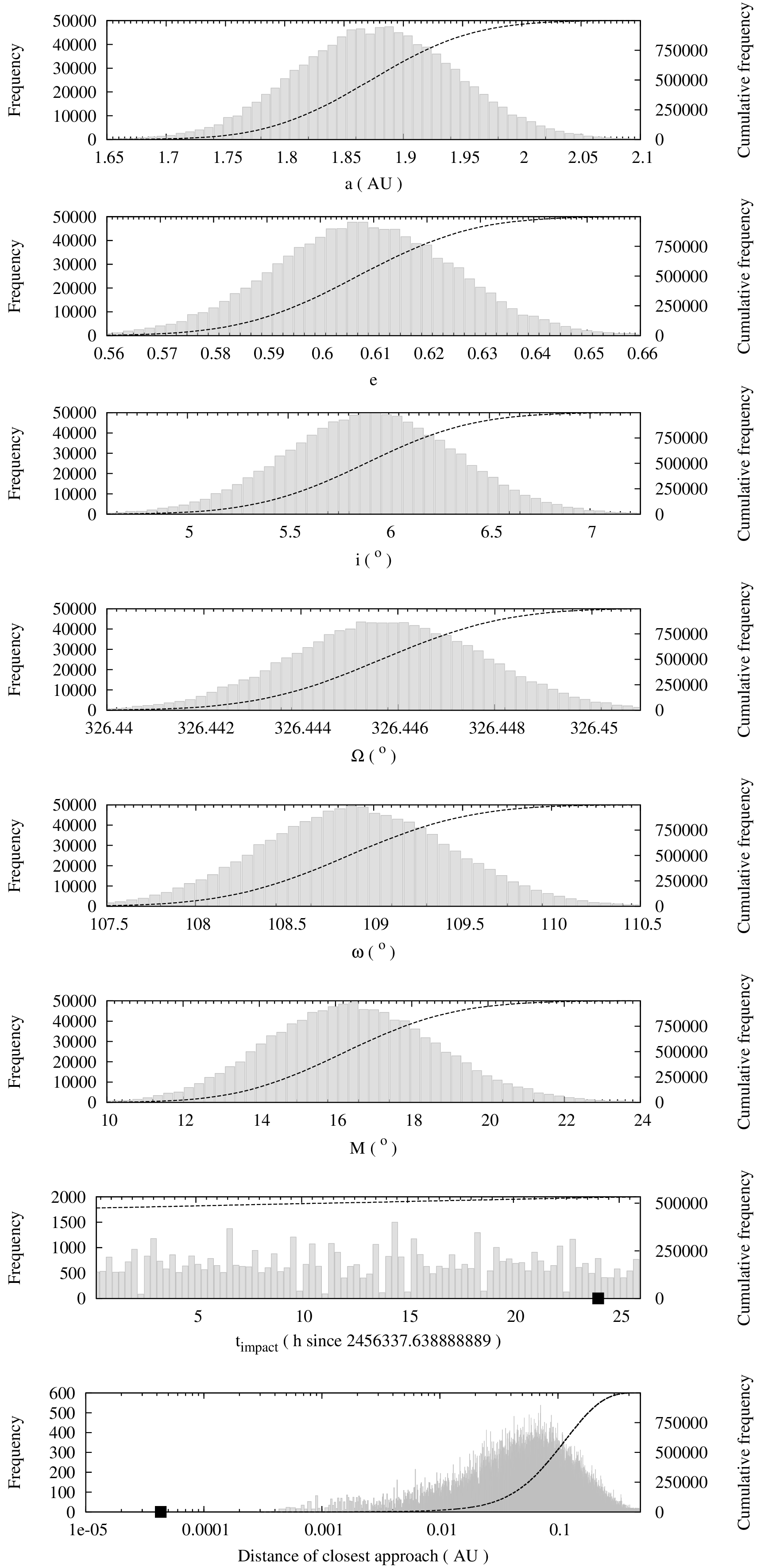}
        \caption{Outcome of an experiment similar to that in Figure \ref{borovicka} but for the orbital solution described in Emel'Yanenko 
                 et al. (2014). A very small fraction of the orbits tested reach their minimal distance to the Earth during the time 
                 interval displayed in the impact time panel.}
        \label{emelyanenko}
     \end{figure}

        The input orbital elements and the results of these simulations are plotted in Figures \ref{borovicka} and \ref{popova}. The 
        solution in Borovi\v{c}ka et al. (2013) is slightly better than the one in Popova et al. (2013) as its impact probability, $P_{\rm 
        imp}$, is $\leq 10^{-5}$, about ten times higher than that of the other one (see Table \ref{orbits}). The value of the impact 
        probability has been obtained in the usual way, dividing the number of relevant events by the total number of test orbits studied. 
        The dispersions in both $t_{\rm impact}$ and minimal approach distance are very wide. Most calculated close approaches take place 
        well after or before $t_{\rm impact}$. No impacts at the right coordinates were recorded in our numerical experiments. The closest 
        virtual impact (found for the one in Borovi\v{c}ka et al. 2013) took place almost 15 minutes after $t_{\rm impact}$ at coordinates 
        (69\fdg4 E, 42\fdg4 N) in Kazakhstan: $a$ = 1.735709580 AU, $e$ = 0.575817480, $i$ = 4\fdg875268780, $\Omega$ = 326\fdg460412000, 
        $\omega$ = 107\fdg626661600, and $\tau$ = 2456293.019210 JDCT (the tabulated precision is intended to facilitate verification). It 
        is perhaps worth mentioning here that a difference in $t_{\rm impact}$ of 15 minutes is far from trivial because it is equivalent to 
        over two Earth diameters in terms of space as our planet travels a distance equal to its own average diameter in nearly 7.1 minutes. 
        The results of this independent statistical test are fully consistent with those presented in Paper II. Applying the same test to 
        the solution in Emel'Yanenko et al. (2014), we found that it is even less satisfactory, statistically speaking (see Figures 
        \ref{popova} and \ref{emelyanenko}), than the one in Popova et al. (2013).
%
%
%
%
%
%
%
%
%
%
%
%

     In Papers I and II, the orbital solutions were derived using a geometric Monte Carlo approach. Such a technique is relatively 
     inexpensive in computational terms and it is able to produce reasonably precise results using very few input data, but the computed 
     solution is not a true pre-impact orbit in the sense explained above. Applying an impact test ---analogous to the one used above--- to 
     the orbital solution presented in Paper II, we obtain $P_{\rm imp} \leq$ 0.77 and the virtual impacts take place 8--15 minutes before 
     $t_{\rm impact}$ with a range in longitude of (31, $-$75)\degr and latitude of (24, 79)\degr. 

  \section{Chasing impactors \textit{N}-body style}
     The full $N$-body statistical test applied in the previous section is certainly robust enough to show that a candidate solution must be
     incorrect. This technique is particularly well suited to be used within the framework of the inverse problem paradigm in which one 
     starts with the results (the impact parameters) and then calculates the cause (the pre-impact orbit of the impactor). In contrast, the 
     approach followed by Sitarski (1998, 1999, 2006) is that of a forward problem: the causes are known (astrometric observations of a 
     putative impactor) and then the results are computed (if an impact is possible, try to find out when and where). In this context, the 
     approach developed in this section can be seen as a generalization of the techniques explored in Sitarski (1998, 1999, 2006); he uses 
     the pre-impact information as input to develop his methodology to predict asteroid impacts, but here our goal is to reconstruct the 
     causal factors (the pre-impact orbit) that triggered the observed impact using the post-impact data as a starting point. In inverse 
     problem parlance, we are facing a nonlinear inverse problem in which we know the data (observed impact parameters) and we look for the 
     best model parameters (the orbit) such as the governing equations (the full $N$-body treatment) ---or forward operator--- provide the 
     optimal relationship between model and data (see, e.g., Press et al. 2007). 

     Most inverse problems are undetermined as the solutions are degenerate, i.e., not unique, with fewer equations than unknowns. It may be 
     argued that the type of inverse problem studied here (going from impact to orbit) cannot be solved because we seek six unknowns (the 
     set of orbital elements) and the post-impact data are just three quantities $t_{\rm impact}$, $\lambda_{\rm impact}$, and $\phi_{\rm 
     impact}$. However, we also have $h_{\rm impact}$, an estimate of the value of the velocity at $h_{\rm impact}$, $v_{\rm impact}$, and 
     the standard deviations of all these quantities. On the other hand, it is a well known fact used in probabilistic curve reconstruction 
     (see, e.g., Unnikrishnan et al. 2006, 2010; Unnikrishnan 2008) that if a curve is smooth, the scatter matrix will be elongated and that 
     its major axis, or principal eigenvector, will approximate the direction of the local tangent. It is not unreasonable to assume that 
     the impact trajectory (the pre-impact orbit of the parent body of the superbolide) is smooth in the neighborhood of the impact point 
     (high in the atmosphere) and therefore that the dispersions in $\lambda_{\rm impact}$, $\phi_{\rm impact}$, and $h_{\rm impact}$ 
     provide an appropriate approximation to the local tangent (and indirectly to the instantaneous value of the velocity and its direction) 
     because the principal eigenvector of the data scatter matrix (that contains the values of the variances) is aligned with the true 
     tangent to the impact curve. In this mathematical context our inverse problem may be viable and a solution could be found. In 
     particular, if the available determination of $v_{\rm impact}$ is robust then the solution of the inverse problem is strictly unique.

     The practical implementation of the solution to this inverse problem requires the use of Monte Carlo techniques (Metropolis \& Ulam 
     1949). As in the previous section, we assume a set of orbital elements ($a$, $e$, $i$, $\Omega$, $\omega$, and $\tau$) at a given epoch 
     $t_0$, generate Cartesian state vectors ($\textit{\textbf{r}}$ and $\textit{\textbf{v}}$) for the assumed orbit at the reference epoch, 
     and use $N$-body simulations within the same physical model applied above to study the evolution of the assumed orbit until an impact 
     or a miss occurs. In order to rank the computed solution ---if it results in a virtual impact--- we use $t_{\rm impact}$, $\lambda_{\rm 
     impact}$, and $\phi_{\rm impact}$, and a trivariate Gaussian distribution:  
     \begin{equation}
        \Psi = {\Large e}^{-\frac{1}{2}\left[
                 \left(\frac{\lambda - \lambda_{\rm impact}}{\sigma_{\lambda_{\rm impact}}}\right)^{2} +
                        \ \left(\frac{\phi - \phi_{\rm impact}}{\sigma_{\phi_{\rm impact}}}\right)^{2} + 
                        \ \left(\frac{t - t_{\rm impact}}{\sigma_{t_{\rm impact}}}\right)^{2} 
               \right]} \,,
               \label{rank}
     \end{equation}
     where $\lambda$ and $\phi$ are the impact coordinates, $t$ is the impact time for the assumed test orbit, and $\sigma_{\lambda_{\rm
     impact}}$, $\sigma_{\phi_{\rm impact}}$, and $\sigma_{t_{\rm impact}}$ are the standard deviations associated with $\lambda_{\rm impact}$, 
     $\phi_{\rm impact}$, and $t_{\rm impact}$, respectively, supplied with the observational impact values. The closer the value of $\Psi$ 
     to 1, the better. The functional form of $\Psi$ assumes that there is no correlation between $\lambda$, $\phi$, and $t$. In our 
     implementation, $\Psi$ is our objective function and our algorithm usually converges after exploring several million orbits. Seeking 
     the optimal orbit can be (and it was) automated using a feedback loop to accelerate convergence in real time. If enough test orbits are 
     studied, the best pre-impact orbit can be determined. This assumption is based on the widely accepted idea that statistical results of 
     an ensemble of collisional $N$-body simulations are accurate, even though individual simulations are not (see, e.g., Boekholt \& 
     Portegies Zwart 2015). 

  \section{Pre-impact orbit}
\begin{figure}
        \centering
        \includegraphics[width=\linewidth]{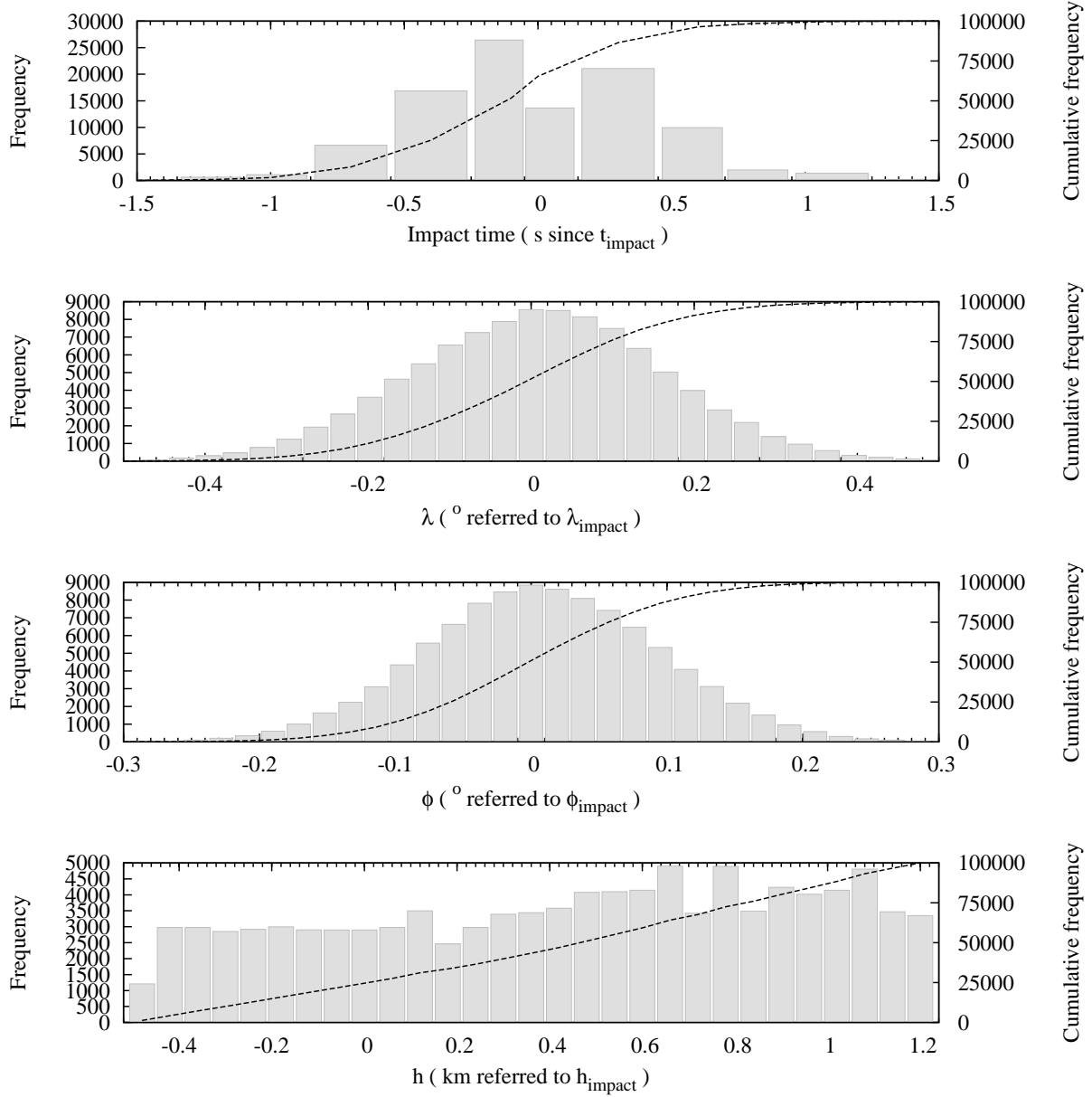}
        \caption{Resulting distribution in impact parameter space ($t$, $\lambda$, $\phi$, and $h$) for an experiment using the orbital 
                 solution in Table \ref{ours} and $10^5$ test orbits. The setup is equivalent to those of Figures 
                 \ref{borovicka}--\ref{emelyanenko}. The values of the virtual impact parameters are referred to those in Popova et al. 
                 (2013) as pointed out above. The rather ragged distribution in impact time is the result of the unavoidable discretization 
                 of the output interval that also has an effect on the distribution in altitude.} 
        \label{testours}
     \end{figure}

     Using $t_{\rm impact}$, $\lambda_{\rm impact}$, and $\phi_{\rm impact}$ to select the best solution and after a few million trials, we 
     obtain the orbital solution (see Tables \ref{orbits} and \ref{ours}): $a$~=~1.62470348 AU, $e$~=~0.53184268, $i$~=~3$\fdg$9749908, 
     $\Omega$~=~326$\fdg$4607324, $\omega$~=~109$\fdg$7012184, and $\tau$~=~2456292.5834112 JDCT, with a value of the geocentric velocity at 
     impact of 17.74 km s$^{-1}$ and an impact probability $>$0.99999. The value of the velocity derived by Miller et al. (2013) from 
     satellite data and video recordings amounts to 17.7$\pm$0.5 km s$^{-1}$. A value of 17.6 km s$^{-1}$ is also favored in Proud (2013). 
     The agreement between our virtual impact parameter results and the observational values, both in terms of impact time and coordinates, 
     is very good (see Table \ref{ours} and Figure \ref{testours}). The velocity at $h_{\rm impact}$ also matches well the one derived by 
     Miller et al. (2013). Therefore, it is a reasonable solution ---statistically speaking--- to the inverse problem pointed out above. 
     This new, most probable orbital solution is not too different from that in Paper II and matches well the one originally computed by S. 
     Nakano (see Table 1 in Paper II).$^5$ 
%
%
     \begin{table}
      \centering
      \fontsize{8}{11pt}\selectfont
      \tabcolsep 0.15truecm
      \caption{Heliocentric Keplerian Orbital Elements of the Chelyabinsk Impactor at Epoch JDCT 2456337.638888889 from our $N$-body 
               Approach  
              }
      \begin{tabular}{lc}
       \hline
        Semimajor axis, $a$ (AU)                           &   1.62470348$\pm$0.00000005   \\
        Eccentricity, $e$                                  &   0.53184268$\pm$0.00000002   \\
        Inclination, $i$ (deg)                             &   3.9749908$\pm$0.0000008     \\
        Longitude of the ascending node, $\Omega$ (deg)    & 326.4607324$\pm$0.0000014     \\
        Argument of perihelion, $\omega$ (deg)             & 109.7012184$\pm$0.0000008     \\
        Mean anomaly, $M$ (deg)                            &  21.4432449$\pm$0.0000013     \\
        Time of perihelion passage, $\tau$ (JDCT)          & 2456292.5834112$\pm$0.0000010 \\
                                                           & 2012 Dec 31 02:00:06.6 UT     \\
        Perihelion, $q$ (AU)                               &   0.760616827$\pm$0.000000012 \\
        Aphelion, $Q$ (AU)                                 &   2.48879014$\pm$0.00000010   \\
        Impact time, $t_{\rm impact}$ (JDCT)               & 2456338.6391296$\pm$0.0000009 \\
                                                           & 2013 Feb 15 03:20:20.8 UT     \\
        Longitude of impact, $\lambda_{\rm impact}$ (deg)  &  64.5649$\pm$0.0014           \\
        Latitude of impact, $\phi_{\rm impact}$ (deg)      & +54.4450$\pm$0.0007           \\
        Altitude of impact, $h_{\rm impact}$ (km)          &  97.9$\pm$0.3                 \\
        Velocity at impact, $v_{\rm impact}$ (km s$^{-1}$) &  17.74110$\pm$0.00013         \\
        Radiant R.A., $\alpha_{0}$ (deg)                   & 334.23104$\pm$0.00005         \\
        Radiant decl., $\delta_{0}$ (deg)                  &  -0.14575$\pm$0.00005         \\
        Radiant velocity, $v_{\rm g}$ (km s$^{-1}$)        &  13.86900$\pm$0.00005         \\
       \hline
      \end{tabular}
      \label{ours}
      \tablenotetext{}{Note. Values include the 1$\sigma$ uncertainty. These values are the result of the average of the 36 best solutions
                       ranked as explained in the text. The impact probability associated with this solution is $>$0.99999. The values of 
                       the velocities quoted here are geocentric and not apparent. In the following this is considered as SOL1.
                      }
     \end{table}
%
%
%
%
%
%

     A problematic issue that compromises the uniqueness of any impact solution is in the uncertainty associated with the value of the 
     geocentric velocity at $h_{\rm impact}$. Proud (2013) already pointed out a range of 17--18.6 km s$^{-1}$ in $v_{\rm impact}$, 
     generating extreme minimum and maximum trajectories. The U.S. Government sensors give a value of 18.6 km s$^{-1}$ (no errors quoted) 
     for the pre-impact geocentric velocity at an altitude of 23.3 km (peak brightness); the value has been obtained by D. Yeomans and P. 
     Chodas.\footnote{http://neo.jpl.nasa.gov/fireball/} Table S1 in Popova et al. (2013) shows that the value of the apparent velocity of 
     the superbolide remained fairly constant between the altitudes of 97.1 and 27 km. On the other hand, in Table 1 of Borovi\v{c}ka et al. 
     (2013) it is indicated that the speed of the Chelyabinsk impactor relative to the Earth's surface high in the atmosphere was 
     19.03$\pm$0.13~km~s$^{-1}$; the value of the apparent velocity at the entry point (97.1$\pm$1.6~km above the ground) of the Chelyabinsk 
     meteoroid given in Table S1 of Popova et al. (2013) is 19.16$\pm$0.15~km~s$^{-1}$. The values of the velocities obtained from our 
     simulations are true geocentric values rather than apparent ones as like those quoted in some of the papers cited. But what is the 
     effect of the uncertainty in $v_{\rm impact}$ on the orbital solution? 

     In an attempt to answer the legitimate concern that prompted this question, we have performed additional calculations using the same 
     data in terms of impact time and coordinates, but changing the model so suitable values of $v_{\rm impact}$ are obtained. We have used
     the same approach described above but this time forcing the model (the orbital parameters) to reach values of $v_{\rm impact}$ equal to 
     $\sim$17 km s$^{-1}$ (SOL0, left-hand column in Table \ref{alt}), $\sim$18.6 km s$^{-1}$ (SOL2, center column in Table \ref{alt}), and 
     $\sim$19.1 km s$^{-1}$ (SOL3, right-hand column in Table \ref{alt}). The impact probabilities associated with these orbital solutions are
     all $>$0.99999. If we compare the orbital solutions in Table \ref{orbits} with those in Table \ref{alt} it is easy to realize why some of
     the published solutions fail to generate any significant impacts. If $v_{\rm impact}$ was $\sim$19.1 km s$^{-1}$ then the closest 
     solution (but still far from satisfactory, see above) is the one in Emel'Yanenko et al. (2014).

     As in the case of SOL1, the agreement between our virtual impact parameter results and the observational values, both in terms of 
     impact time and coordinates, is very good and their impact probabilities essentially equal to 1. However, the $v_{\rm impact}$ and the 
     parameters of the true radiant (see below) of these alternative solutions are quite different. Fortunately, the orbital elements follow 
     a linear relationship with $v_{\rm impact}$ and if a truly robust value of this parameter is eventually found (for example, from still 
     unreleased military radar data), the following equations can be used to determine a first approximation for the appropriate orbit ($a$ 
     in AU, $i$, $\Omega$, and $\omega$ in degrees, $\tau$ in JDCT, and $v_{\rm impact}$ in km s$^{-1}$):
     \begin{eqnarray}
        a      & = & (0.144\pm0.005)\ v_{\rm impact} - (0.93\pm0.09)\,, r^2 = 0.9979\,, \\               
        e      & = & (0.0476\pm0.0005)\ v_{\rm impact} - (0.313\pm0.009)\,, r^2 = 0.9998\,, \\               
        i      & = & (0.86\pm0.03)\ v_{\rm impact} - (11.3\pm0.6)\,, r^2 = 0.9970\,, \\               
        \Omega & = & (-0.0082\pm0.0005)\ v_{\rm impact} + (326.606\pm0.008)\,, r^2 = 0.9935\,, \\               
        \omega & = & (-0.342\pm0.011)\ v_{\rm impact} + (115.8\pm0.2)\,, r^2 = 0.9978\,, \\               
        \tau   & = & (1.17\pm0.03)\ v_{\rm impact} + (2456271.8\pm0.5)\,, r^2 = 0.9989\,,                
     \end{eqnarray}
     These expressions have been obtained from SOL0 to SOL3 and their correlation coefficients, or Pearson's $r$ (see, e.g., Press et al. 
     2007), are very good. However, the precision of the values obtained from these equations is limited by the errors associated with the 
     linear regression parameters (one to three decimal places, depending on the orbital element). These expressions give relatively 
     low-precision estimates for the values of the orbital elements and may not apply outside the $v_{\rm impact}$ range of (17, 19.1) km 
     s$^{-1}$. 
%
%
     \begin{landscape}
     \begin{table}
      \centering
      \fontsize{8}{11pt}\selectfont
      \tabcolsep 0.15truecm
      \caption{Alternative Heliocentric Keplerian Orbital Elements of the Chelyabinsk Impactor at Epoch JDCT 2456337.638888889 from our 
               $N$-body Approach.  
              }
      \begin{tabular}{lccc}
       \hline
        Semimajor axis, $a$ (AU)                           &   1.53892950$\pm$0.00000004   &   1.74875174$\pm$0.00000003   &   1.83680075$\pm$0.00000005   \\
        Eccentricity, $e$                                  &   0.49874206$\pm$0.00000002   &   0.572221340$\pm$0.000000009 &   0.596987671$\pm$0.000000015 \\
        Inclination, $i$ (deg)                             &   3.4927759$\pm$0.0000005     &   4.7550334$\pm$0.0000005     &   5.2563565$\pm$0.0000007     \\
        Longitude of the ascending node, $\Omega$ (deg)    & 326.4672008$\pm$0.0000008     & 326.4537788$\pm$0.0000007     & 326.4504518$\pm$0.0000010     \\
        Argument of perihelion, $\omega$ (deg)             & 109.9462609$\pm$0.0000009     & 109.4416222$\pm$0.0000005     & 109.2257672$\pm$0.0000007     \\
        Mean anomaly, $M$ (deg)                            &  23.6960400$\pm$0.0000013     &  18.7763467$\pm$0.0000006     &  17.2169040$\pm$0.0000009     \\
        Time of perihelion passage, $\tau$ (JDCT)          & 2456291.7402523$\pm$0.0000007 & 2456293.5834773$\pm$0.0000004 & 2456294.1534305$\pm$0.0000006 \\
                                                           & 2012 Dec 30 05:45:57.8 UT     & 2013 Jan 01 02:00:12.4 UT     & 2013 Jan 01 15:40:56.4 UT     \\
        Perihelion, $q$ (AU)                               &   0.771400634$\pm$0.000000012 &   0.748078675$\pm$0.000000010 &   0.740253347$\pm$0.000000012 \\
        Aphelion, $Q$ (AU)                                 &   2.30645836$\pm$0.00000008   &   2.74942480$\pm$0.00000006   &   2.93334814$\pm$0.00000011   \\
        Impact time, $t_{\rm impact}$ (JDCT)               & 2456338.6391287$\pm$0.0000005 & 2456338.6391305$\pm$0.0000005 & 2456338.6391287$\pm$0.0000005 \\
                                                           & 2013 Feb 15 03:20:20.7 UT     & 2013 Feb 15 03:20:20.9 UT     & 2013 Feb 15 03:20:20.7 UT     \\
        Longitude of impact, $\lambda_{\rm impact}$ (deg)  &  64.566$\pm$0.008             &  64.567$\pm$0.008             &  64.567$\pm$0.009             \\
        Latitude of impact, $\phi_{\rm impact}$ (deg)      & +54.441$\pm$0.004             & +54.447$\pm$0.005             & +54.443$\pm$0.005             \\
        Altitude of impact, $h_{\rm impact}$ (km)          &  96.87$\pm$0.11               &  98.16$\pm$0.13               &  96.8$\pm$0.2                 \\
        Velocity at impact, $v_{\rm impact}$ (km s$^{-1}$) &  17.06933$\pm$0.00007         &  18.59238$\pm$0.00007         &  19.13400$\pm$0.00010         \\
        Radiant R.A., $\alpha_{0}$ (deg)                   & 334.58835$\pm$0.00005         & 333.65322$\pm$0.00004         & 333.26714$\pm$0.00004         \\
        Radiant decl., $\delta_{0}$ (deg)                  &  $-$0.65857$\pm$0.00005       &  +0.786599$\pm$0.000009       &  +1.31902$\pm$0.00004         \\
        Radiant velocity, $v_{\rm g}$ (km s$^{-1}$)        &  12.99800$\pm$0.00005         &  14.94270$\pm$0.00005         &  15.61070$\pm$0.00005         \\
       \hline
      \end{tabular}
      \label{alt}
      \tablenotetext{}{Note. Values include the 1$\sigma$ uncertainty. The impact probabilities associated with these solutions are 
                       $>$0.99999. The values of the velocities quoted here are geocentric and not apparent. In the following these 
                       solutions are referred to as SOL0 (left-hand column), SOL2 (center column), and SOL3 (right-hand column).
                      }
     \end{table}
     \end{landscape}
%
%

\begin{figure}
        \centering
        \includegraphics[width=0.8\linewidth]{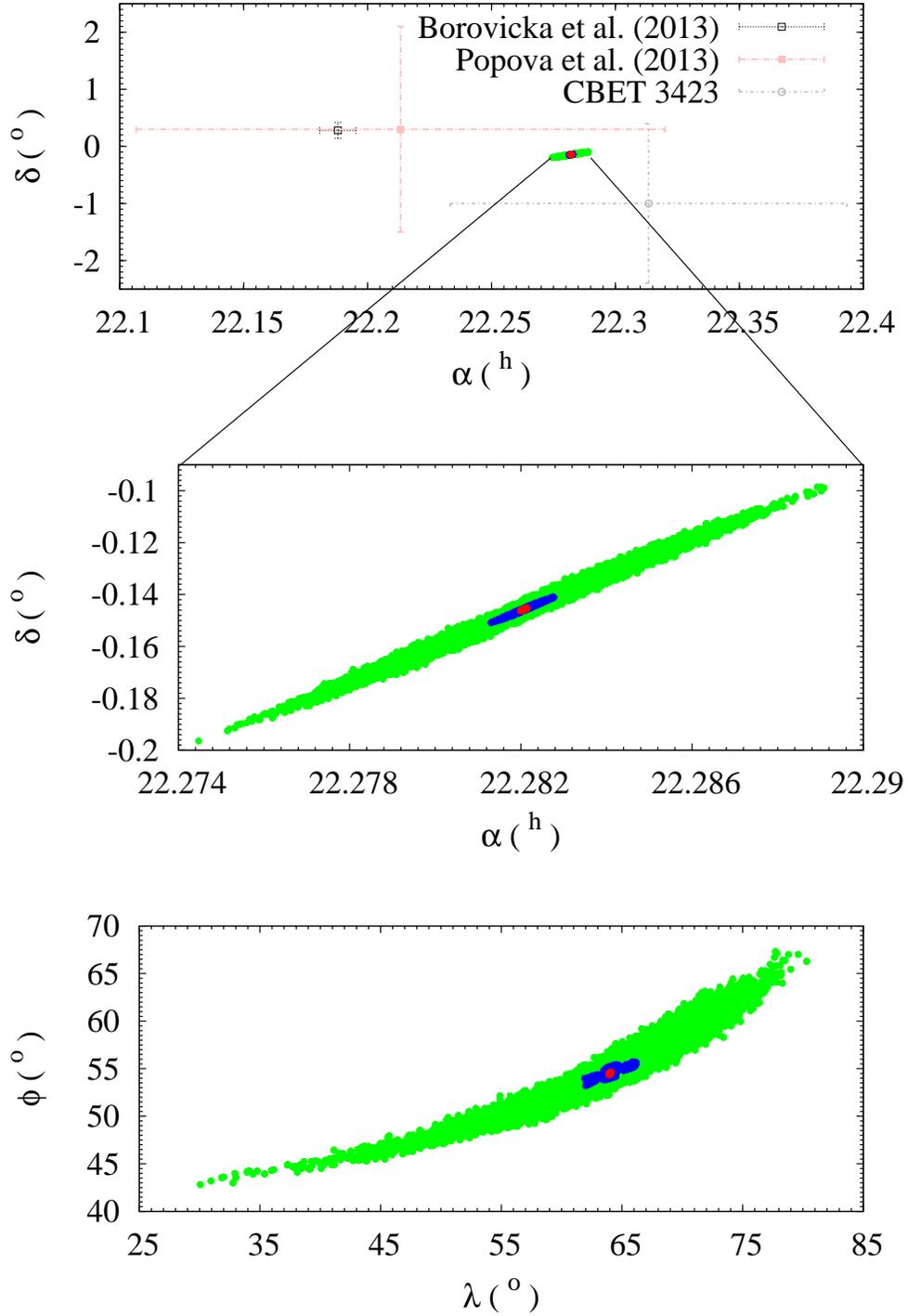}
        \caption{Geocentric equatorial coordinates of the radiant (top and middle panels) and their associated virtual impact coordinates 
                 (bottom panel). Virtual impacts plotted in green represent those associated with sets of orbital elements within 100$\sigma$ 
                 of those in Table \ref{ours}, the ones in blue are the result of a 10$\sigma$ spread, and those in red are restricted to 
                 1$\sigma$ (see the text for further details). 
                }
        \label{radiant}
     \end{figure}

     Another important observational parameter associated with an impact event is the radiant or point in the sky from which the incoming 
     meteor appeared to originate. Figure \ref{radiant}, top panel, shows the location in geocentric equatorial coordinates of the radiant
     point as initially computed by Borovi\v{c}ka et al. in CBET 3423 (Green 2013; gray point and error bars), Borovi\v{c}ka et al. (2013; 
     black point and error bars), and Popova et al. (2013; pink point and error bars). Aiming at comparing with these observational 
     determinations, we have computed the true radiant geocentric equatorial coordinates associated with the orbital solution in Table 
     \ref{ours}. In order to understand better the effect of uncertainties in the computation of both radiants and impact points we have
     performed three sets of simulations with $2\times10^4$ test orbits each: the first set has been generated within 1$\sigma$ of the 
     values in Table \ref{ours}, the second set corresponds to a 10$\sigma$ spread, and the third one has a 100$\sigma$ dispersion. The test 
     orbits have been computed using uniformly distributed random numbers ---not Gaussian like in the rest of this work--- in order to 
     survey the relevant volume of the orbital parameter space evenly. The geocentric equatorial coordinates resulting from our simulations 
     are true values. In contrast, observational determinations give us the position of the incoming object when its light left the 
     impactor. These apparent values have to be corrected for this time delay to obtain the true position of the object in the sky when it 
     was observed. No comments are made in Green (2013), Borovi\v{c}ka et al. (2013), or Popova et al. (2013) regarding possible 
     corrections; however, the error bars in Green (2013) or Popova et al. (2013) are so large that any correction made is probably 
     irrelevant. The arc in coordinates of the geocentric radiant described by SOL0 to SOL3 goes from (22\fh306, $-$0\fdg659) to (22\fh218, 
     +1\fdg319) and its geocentric velocity, $v_{\rm g}$, ranges from 12.998 to 15.611 km s$^{-1}$. The value of $v_{\rm g}$ in 
     Borovi\v{c}ka et al. (2013) is 15.14$\pm$0.16 km s$^{-1}$; Popova et al. (2013) gives a value of 15.3$\pm$0.4 km s$^{-1}$. The 
     approximate values of the coordinates of the geocentric radiant, $\alpha_{0}$ and $\delta_{0}$, and $v_{\rm g}$, as a function of 
     $v_{\rm impact}$ for solutions SOL0 to SOL3 are given by the expressions ($\alpha_{0}$ and $\delta_{0}$ in degrees, and $v_{\rm g}$ and 
     $v_{\rm impact}$ in km s$^{-1}$):
     \begin{eqnarray}
        \alpha_{0} & = & (-0.64\pm0.03)\ v_{\rm impact} + (345.6\pm0.5)\,, r^2 = 0.9965\,, \\               
        \delta_{0} & = & (0.98\pm0.05)\ v_{\rm impact} - (17.4\pm0.9)\,, r^2 = 0.9953\,, \\               
        v_{\rm g}  & = & (1.265\pm0.009)\ v_{\rm impact} - (8.6\pm0.2)\,, r^2 = 0.9999\,.                
     \end{eqnarray}

\begin{figure}
        \centering
        \includegraphics[width=\linewidth]{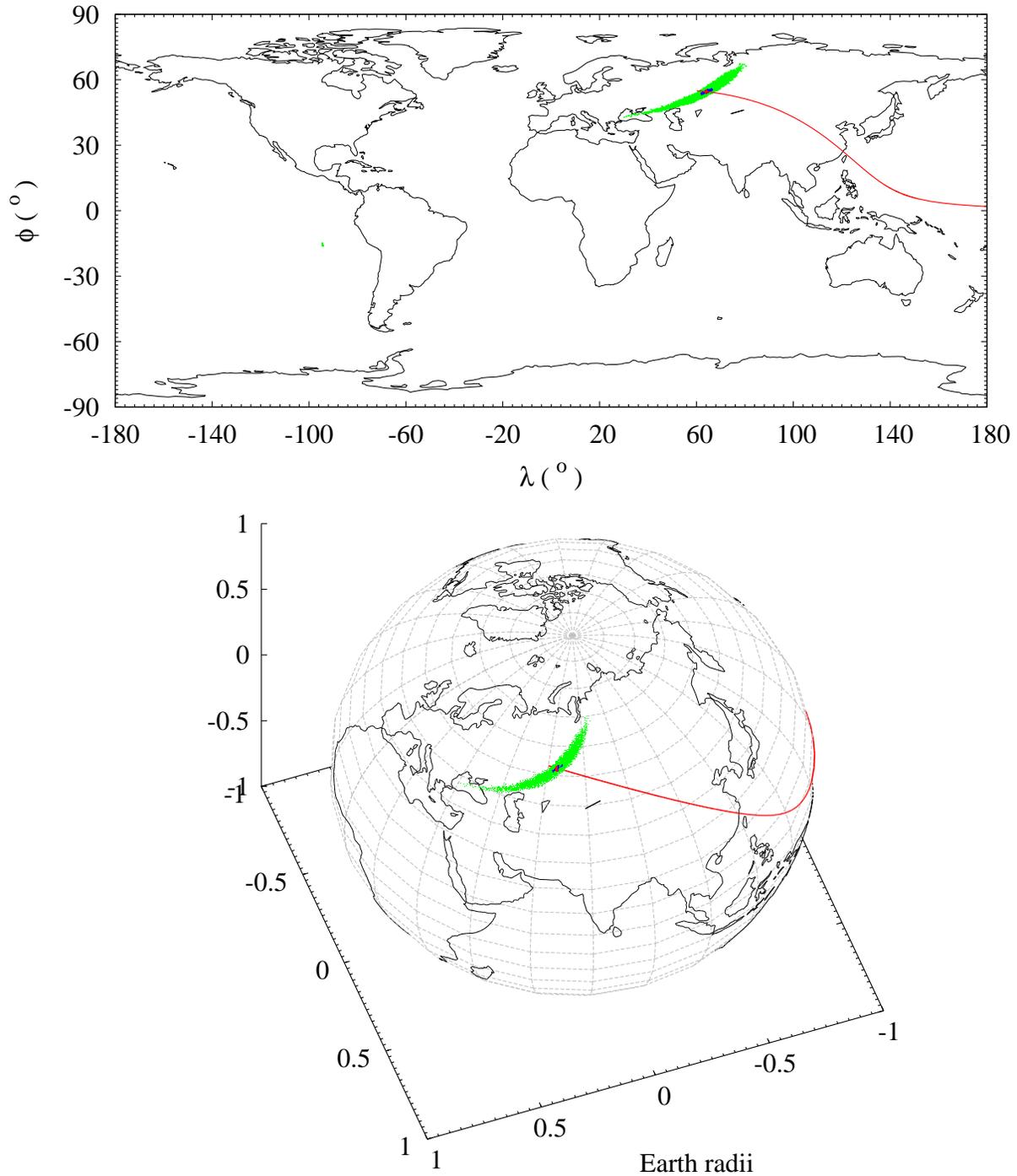}
        \caption{Distribution on the surface of the Earth of the virtual impacts studied in Figure \ref{radiant}, using the same color 
                 coding. The virtual impacts define an arc extending from the Black Sea to the Siberian Plain if deviations as high as 
                 100$\sigma$ are considered. The projected flight path of the nominal solution in Table \ref{ours} is plotted as a red 
                 curve. 
                }
        \label{sphere}
     \end{figure}

\begin{figure}
        \centering
        \includegraphics[width=\linewidth]{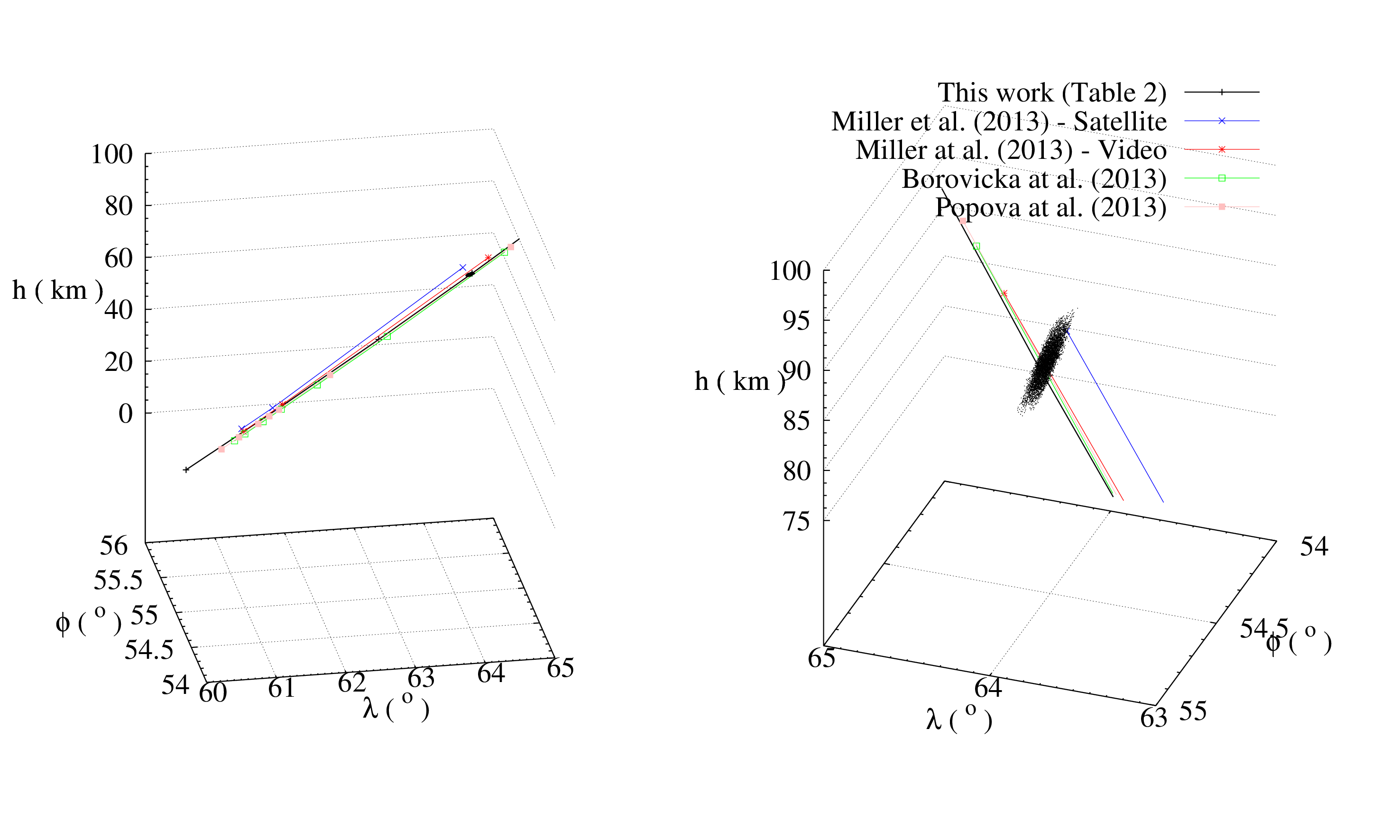}
        \caption{Terminal phase of the impact trajectory of the Chelyabinsk superbolide as described by the solution in Table \ref{ours}, 
                 the two determinations shown in Figure 5 in Miller et al. (2013), and those in Table 1 in Borovi\v{c}ka et al. (2013) and 
                 Table S1 in Popova et al. (2013). The virtual impacts plotted as red points in Figures \ref{radiant} and \ref{sphere} are 
                 replotted here as black dots. The right-hand figure is a rotated and magnified version of the left-hand figure.
                }
        \label{miller}
     \end{figure}

     The colored spot in Figure \ref{radiant}, top panel, shows the true position of the radiant associated with the orbital solution in Table 
     \ref{ours} (SOL1). A magnified version of that area is displayed in the middle panel of Figure \ref{radiant}. Here, the panel shows the 
     true geocentric equatorial coordinates of the virtual impactors at the beginning of the simulation, i.e., at epoch 2456337.638888889 
     JDCT. The points in red correspond to the set of test orbits with orbital elements within 1$\sigma$ of the solution in Table \ref{ours}, 
     those in blue have a 10$\sigma$ spread, and the green ones have 100$\sigma$. Each virtual impactor generates one point on the bottom 
     panel of Figure \ref{radiant} following the same color pattern. The distribution on the surface of the Earth of the virtual impacts 
     studied in Figure \ref{radiant} is better visualized in Figure \ref{sphere} where the virtual impacts define an arc extending from the 
     Black Sea to the Siberian Plain if deviations as high as 100$\sigma$ are allowed in the initial conditions. The separation in impact 
     time between the two most extreme test orbits in the 100$\sigma$ set is nearly 3.5 minutes. The projected flight path of the nominal 
     solution in Table \ref{ours} is plotted as a red curve. This figure is similar to panel (b), Figure 5 in Sitarski (1998). Figure 
     \ref{radiant}, bottom panel, gives a very clear idea of how precise an orbital solution must be in order to make reliable predictions 
     regarding the location of a future strike once a candidate impactor has been identified. The direction of flight in Figure \ref{sphere} 
     matches well that in Figure 4 in Miller et al. (2013). Figure \ref{miller} shows a comparison between our results and those shown in 
     Figure 5 in Miller et al. (2013). The satellite-derived (blue) and surface-based video (red) reconstructions of the impact trajectory 
     of the Chelyabinsk superbolide presented by Miller et al. (2013) are part of the 3$\sigma$ sample associated with the solution in Table 
     \ref{ours}. 
%
%
%
%
%
%
%
%
%
%
%
%

     The values of the coordinates of the geocentric radiant are $\alpha_{0}=334\fdg23104\pm0\fdg00005$ (or $22\fh282069\pm0\fh000003$) and
     $\delta_{0}=-0\fdg14575\pm0\fdg00005$; the geocentric velocity associated with the radiant is $v_{\rm g}=13.86900\pm0.00005$ km s$^{-1}$. 
     Apparently, there is a documented meteor stream that may be associated with this radiant. Terentjeva \& Bakanas (2013) have pointed out 
     that the Daytime Pegasids-Aquariids could be the source of the Chelyabinsk impactor. The parameters in their Table 1 marginally match 
     those in our Table \ref{ours} although no estimates of the values of the errors are given in their work. This meteor shower is not 
     documented in the extensive list compiled by Jenniskens (2006), however. Given the tentative link pointed out in Paper II between 
     Chelyabinsk and other LL5 chondrite falls (see Section 5 in Paper II), this finding just adds another piece to this fascinating puzzle.

  \section{Related objects and dynamical evolution}
\begin{figure}
       \centering
        \includegraphics[width=\linewidth]{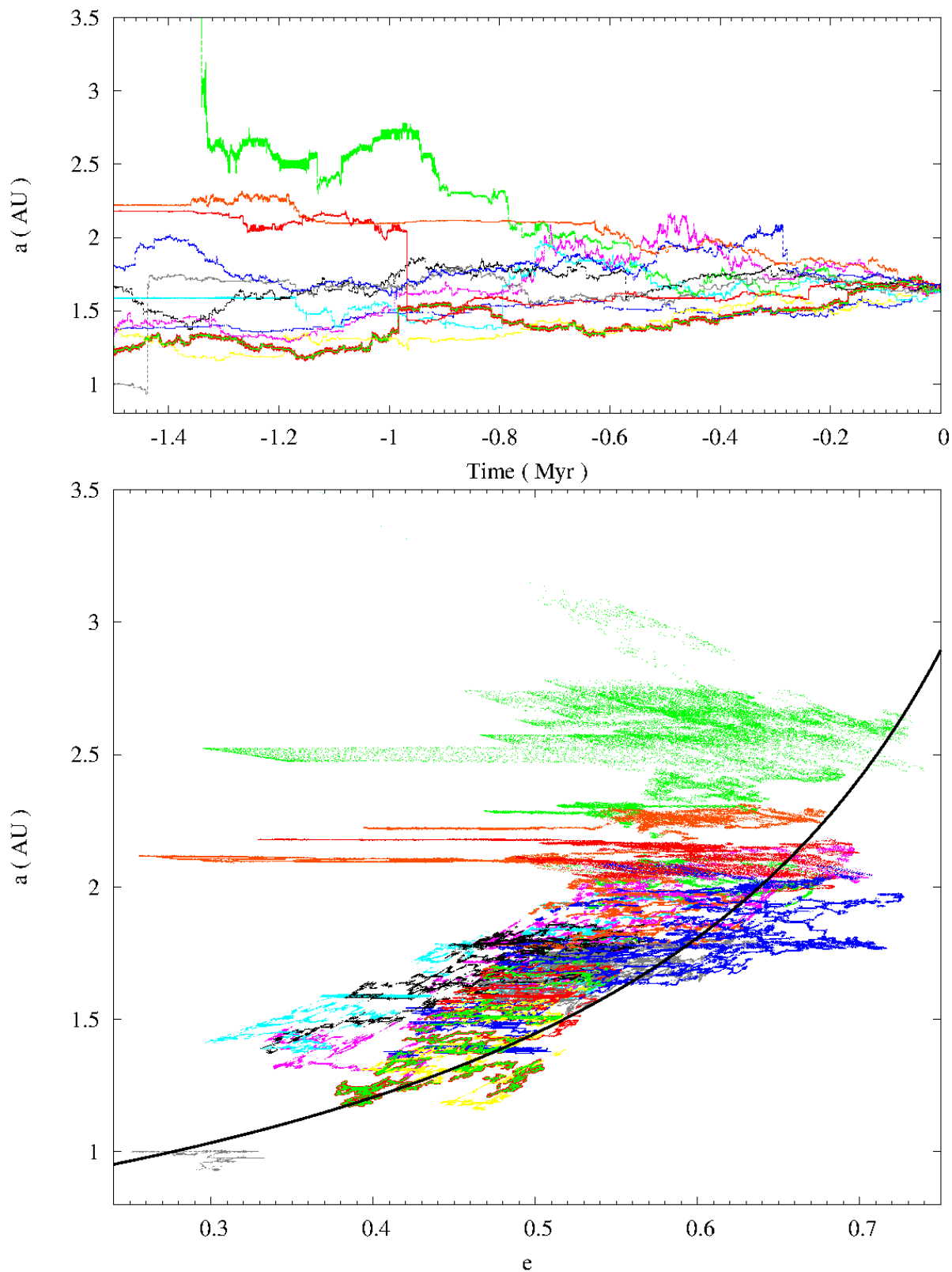}
        \caption{Time evolution of the orbital element $a$ of multiple control orbits of the Chelyabinsk impactor as described by the 
                 solution displayed in Tables \ref{orbits} and \ref{ours} (top panel), and evolution of the same orbits in the $e$--$a$ 
                 plane (bottom panel). Eleven control orbits are plotted; the nominal one is the thick red/green curve. The continuous
                 line represents the ($e$, $a$) combination with perihelion at the semimajor axis of Venus (0.7233 AU). 
                }
        \label{control}
     \end{figure}

     The cosmic-ray exposure age of the Chelyabinsk meteoritic samples has been determined to be about 1.2 Myr (see, e.g., Popova et al. 
     2013). This relatively young age can be interpreted as the approximate time elapsed since the surface of the impactor was first exposed
     to cosmic rays, probably as a result of a break-up event. Figure \ref{control} shows the results of the backwards integration of eleven
     control orbits plus the nominal one in an attempt to explore the probable location of the Chelyabinsk asteroid 1.2 Myr ago, according 
     to the orbital solution in Table \ref{ours}. Our full $N$-body reconstruction of the pre-impact orbit of the Chelyabinsk impactor in 
     Figure \ref{control} places this object directly in the region from 1.2 to 2.8 AU at that time. About 36\% of the orbits have values of 
     the semimajor axis below that of Mars. Another 36\% have $a$ around 1.7 AU. The rest are trapped inside the secular resonance $\nu_{6}$ 
     and jumping into the strong 4:1 mean motion resonance with Jupiter at 2.064 AU as described in Scholl \& Froeschl\'e (1991) or even at 
     the 3:1 orbital resonance with Jupiter (at 2.5 AU) as described by, e.g., Gladman et al. (1997). The figure also shows that the control 
     orbits experience multiple episodes of horizontal (resonant) oscillations. In any case, and if the Chelyabinsk impactor was formed 
     during a fragmentation event nearly 1.2 Myr ago, it is virtually impossible that any related fragments could still be moving in orbits 
     very similar to that in Table \ref{ours}. However, if the path followed by the impactor during the last 1 Myr or so is regarded as a 
     delivery route as described in Morbidelli et al. (1994), it is perfectly possible that other, physically unrelated minor bodies could 
     be following orbits similar to that of the Chelyabinsk impactor, forming a dynamical or resonant group.
%
%
%
%

\begin{figure}
       \centering
        \includegraphics[width=8cm]{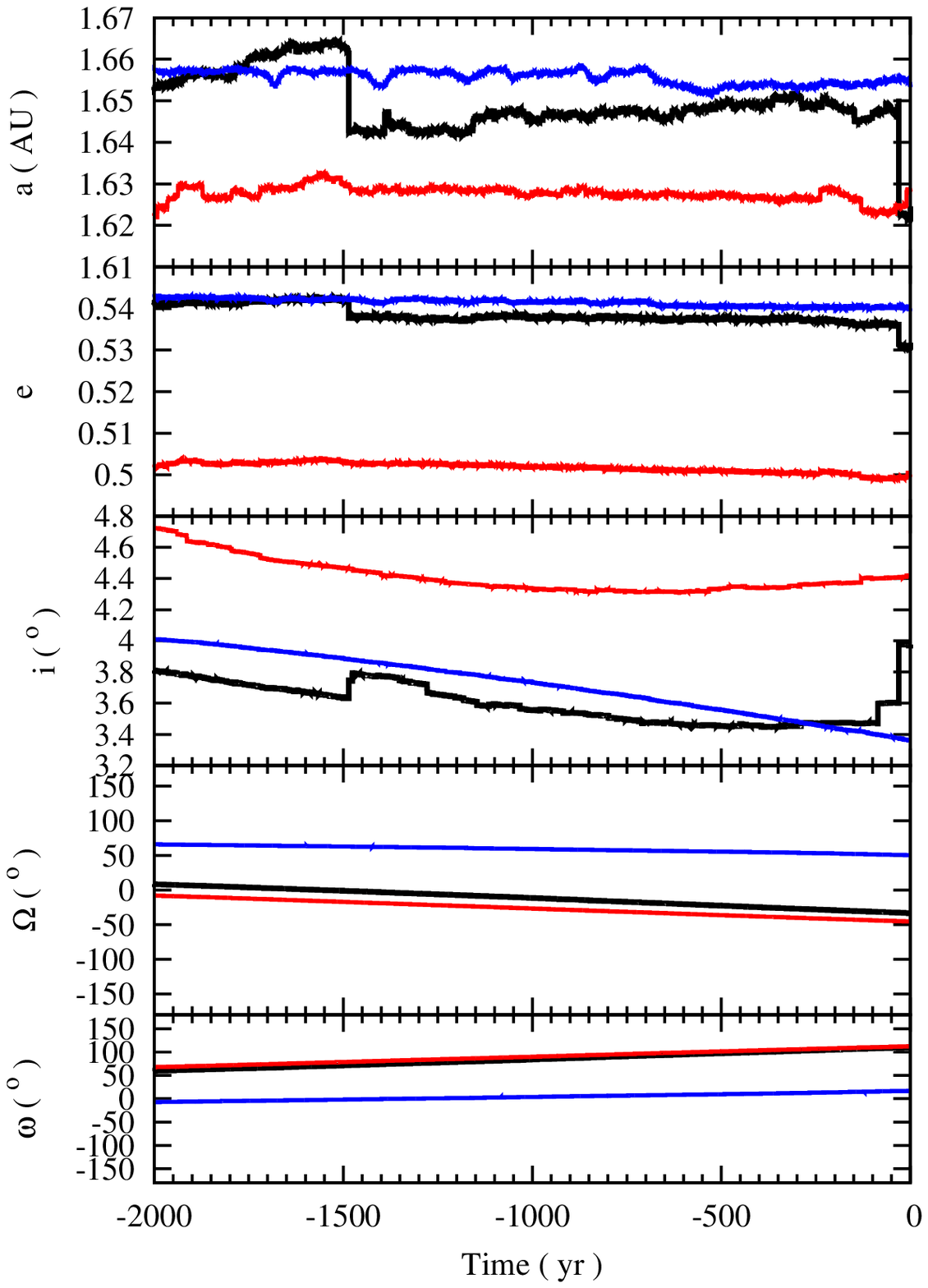}
        \includegraphics[width=8cm]{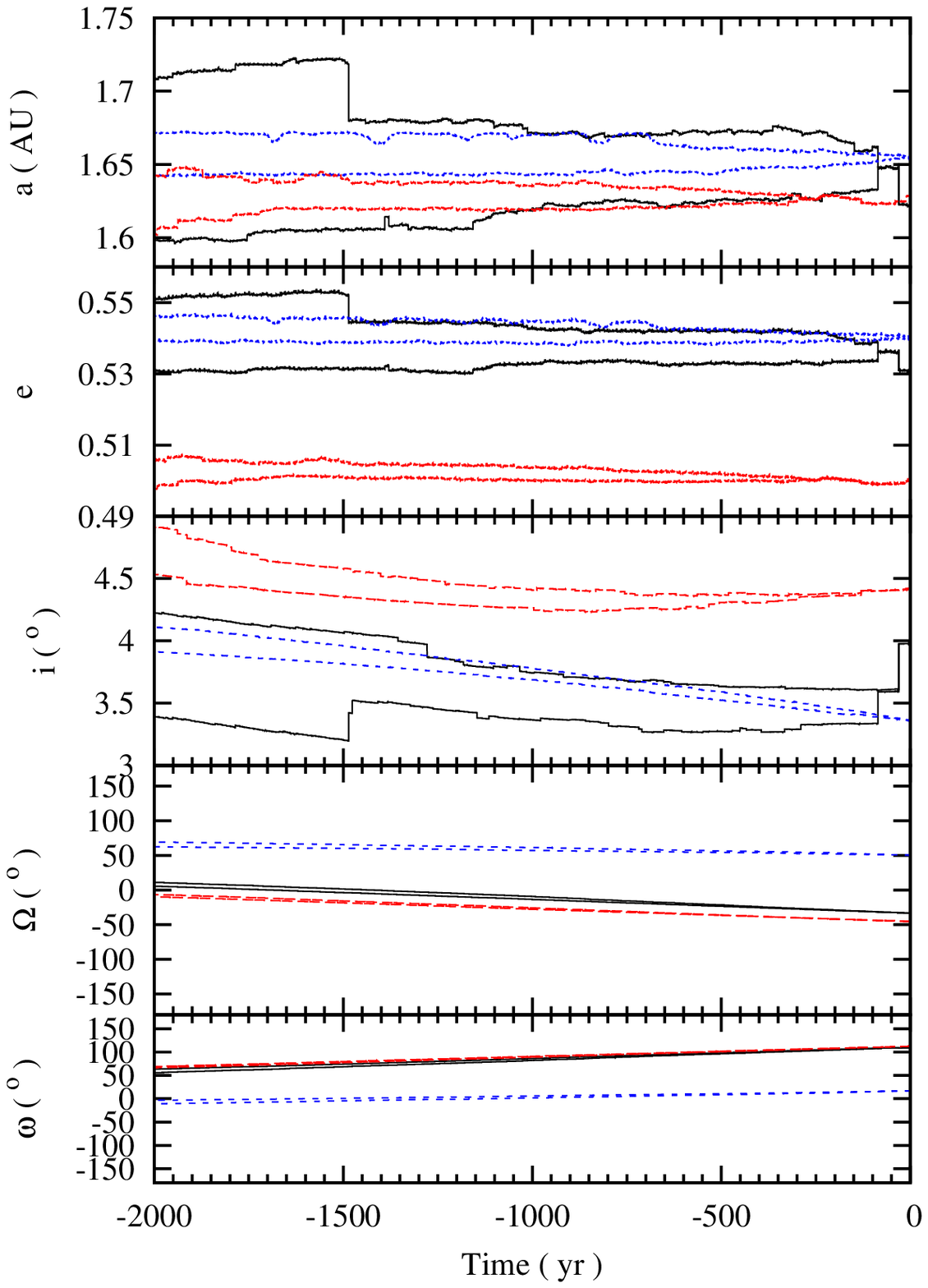}\\
        \caption{Time evolution of the orbital elements $a$, $e$, $i$, $\Omega$, and $\omega$ of 2003~BR$_{47}$ (red), 2011~EO$_{40}$ 
                 (blue), and the Chelyabinsk impactor (black) as described by the solution displayed in Tables \ref{orbits} and \ref{ours}. 
                 The left-hand panels show the average evolution of 100 control orbits, the right-hand panels show the ranges in the values 
                 of the parameters at the given time.
                }
        \label{compare}
     \end{figure}

\begin{figure}
        \centering
        \includegraphics[width=\linewidth]{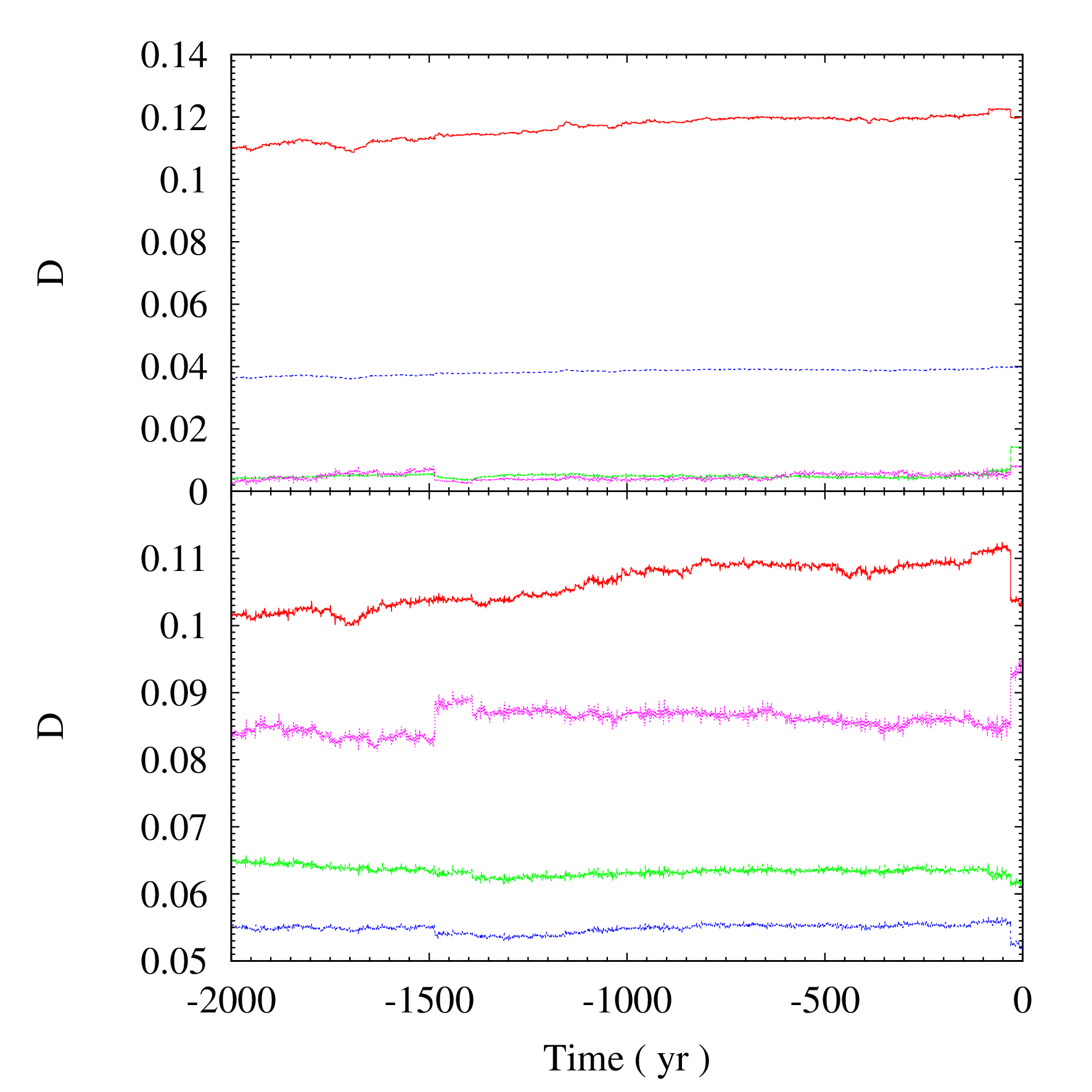}
         \caption{Average time evolution of the various $D$-criteria ---$D_{\rm SH}$ (red), $D_{\rm LS}$ (green), $D_{\rm D}$ (blue), and 
                  $D_{\rm R}$ (pink)--- for 2011~EO$_{40}$ (top panel) and 2003~BR$_{47}$ (bottom panel) with respect to the Chelyabinsk 
                  impactor as described by the solution displayed in Tables \ref{orbits} and \ref{ours}. The values have been computed using 
                  the data in Figure \ref{compare}, left-hand panels.
                 }
         \label{Ds}
      \end{figure}

     Assuming that other objects could be moving in similar orbits, we use the $D$-criteria of Southworth \& Hawkins (1963, $D_{\rm SH}$), 
     Lindblad \& Southworth (1971, $D_{\rm LS}$), Drummond (1981, $D_{\rm D}$), and the $D_{\rm R}$ from Valsecchi et al. (1999) to 
     investigate possible dynamical connections between this object and known minor bodies. A search among all the objects currently 
     catalogued (as of 2015 August 19) by the JPL Small-Body Database\footnote{http://ssd.jpl.nasa.gov/sbdb.cgi} using these criteria gives 
     the list of candidates in Table \ref{candidatesCH}. With a few exceptions, their orbits are poorly constrained as they are based on 
     short arcs but they are provided here to encourage further observations. All of them are classified as Apollos, near-Earth asteroids 
     (NEAs) and, a few, as potentially hazardous asteroids (PHAs); their aphelia are in or near the 3:1 orbital resonance with Jupiter (at 
     2.5 AU). These objects are strongly perturbed as they experience periodic close encounters not only with the Earth--Moon system but 
     also with Mars, Ceres and, in some cases, Venus. They are also subjected to multiple secular resonances (see Paper II). Figure 
     \ref{compare} compares the evolution of the orbital parameters of the Chelyabinsk impactor as described by the solution displayed in 
     Tables \ref{orbits} and \ref{ours} and 2011~EO$_{40}$ that is the known NEO with the lowest values of the various $D$-criteria. Also 
     plotted is the evolution of 2003~BR$_{47}$, which is the object (also NEO and PHA like 2011~EO$_{40}$) with the closest orbit to that 
     of the impactor ---as characterized in Table \ref{ours}--- if the five orbital elements are considered ($a$~=~1.6283325 AU, 
     $e$~=~0.5001041, $i$~=~4$\fdg$42080, $\Omega$~=~314$\fdg$56875, and $\omega$~=~112$\fdg$52038, nominal precision with 
     $D_{\rm SH}$~=~0.1037,  $D_{\rm LS}$~=~0.0626, $D_{\rm D}$~=~0.0529, and $D_{\rm R}$~=~0.0948). Data in Figure \ref{compare} are the 
     result of averaging 100 control orbits generated after applying a Monte Carlo approach within the orbital parameter domain limited by 
     the available 1$\sigma$ uncertainties (see Table \ref{datacandidates}) and using Gaussian random numbers. The average time evolution 
     of the various $D$-criteria for 2003~BR$_{47}$ and 2011~EO$_{40}$ is displayed in Figure \ref{Ds}; this comparison is customarily used 
     to link meteors and NEOs (see, e.g., Trigo-Rodr\'{\i}guez et al. 2007; Olech et al. 2015). Statistically speaking, the Chelyabinsk 
     impactor is a robust dynamical relative of 2011~EO$_{40}$ as described by SOL1: the ranges of their orbital parameters, $a$, $e$, and 
     $i$, fully overlap after about 100 years of backwards integration. Candidate dynamical relatives for SOL0, SOL2, and SOL3 are compiled 
     in Tables \ref{sol0}--\ref{sol3}, respectively.
%
%
     \begin{landscape}
     \begin{table}
      \centering
      \fontsize{8}{11pt}\selectfont
      \tabcolsep 0.07truecm
      \caption{Orbital Elements, Orbital Periods ($P_{\rm orb}$), Perihelia ($q = a \ (1 - e)$), Aphelia ($Q = a \ (1 + e)$), Number of 
               Observations ($n$), Data arc, and Absolute Magnitudes ($H$) of Objects Moving in Orbits Similar to that of the Meteoroid that 
               Caused the Chelyabinsk Superbolide as Described by the Solution Displayed in Tables \ref{orbits} and \ref{ours}} 
      \begin{tabular}{lccccccccccccccccc}
       \hline
          Asteroid       & Epoch & $a$ (AU)  & $e$        & $i$ (deg) & $\Omega$ (deg) & $\omega$ (deg) & $P_{\rm orb}$ (year) & $q$ (AU) & $Q$ (AU) 
                         & $n$ & arc (d) & $H$ (mag) 
                         & $D_{\rm SH}$ & $D_{\rm LS}$ & $D_{\rm D}$ & $D_{\rm R}$ & PHA \\
       \hline
         2011 EO$_{40}$  & 57000 & 1.6540887 & 0.54016233 & 3.36286   &  50.30284      &  17.07473      & 2.13                 & 0.76     & 2.55
                         & 20  & 34      & 21.50 
                         & 0.1198       & 0.0135       & 0.0396      & 0.0073      & Yes \\
         2002 AC$_{9}$   & 57000 & 1.7038247 & 0.56051693 & 2.28428   &   2.58887      &  28.42304      & 2.22                 & 0.75     & 2.66
                         & 51  & 3132    & 21.00 
                         & 0.4232       & 0.0428       & 0.1407      & 0.0224      & Yes \\
         2014 SJ$_{142}$ & 57000 & 1.5591764 & 0.50812802 & 2.08620   & 336.63875      & 119.15404      & 1.95                 & 0.77     & 2.35
                         & 34  & 30      & 23.20 
                         & 0.1821       & 0.0411       & 0.0623      & 0.0234      & No  \\
         2012 QZ$_{16}$  & 57000 & 1.5376728 & 0.50328326 & 6.11985   & 151.62956      & 258.93607      & 1.91                 & 0.76     & 2.31
                         & 23  & 2       & 25.50 
                         & 0.2902       & 0.0472       & 0.1007      & 0.0237      & No  \\
         2015 CC$_{1}$   & 57070 & 1.5467360 & 0.51434097 & 3.59953   &  88.60772      & 327.84872      & 1.92                 & 0.75     & 2.34
                         & 22  & 16      & 23.88 
                         & 0.2149       & 0.0209       & 0.0722      & 0.0248      & No  \\
         2012 VA$_{20}$  & 57000 & 1.6822360 & 0.55511183 & 4.39092   &  62.74253      & 240.11893      & 2.18                 & 0.75     & 2.62
                         & 16  & 10      & 22.80 
                         & 1.0048       & 0.0273       & 0.4053      & 0.0257      & No  \\
         2013 BR$_{15}$  & 56309 & 1.5546110 & 0.52043752 & 1.95468   & 102.89879      & 284.89340      & 1.94                 & 0.75     & 2.36
                         & 10  & 2       & 25.00 
                         & 0.4427       & 0.0400       & 0.1464      & 0.0260      & No  \\
         2009 SD         & 55091 & 1.7343896 & 0.56682933 & 3.04789   & 344.32550      & 287.05721      & 2.28                 & 0.75     & 2.72
                         & 24  & 3       & 25.40 
                         & 1.0899       & 0.0397       & 0.5041      & 0.0265      & No  \\
         1996 VB$_{3}$   & 50401 & 1.6259062 & 0.54444404 & 2.79583   & 180.62271      & 132.68222      & 2.07                 & 0.74     & 2.51
                         & 21  & 9       & 22.40 
                         & 0.9520       & 0.0313       & 0.3687      & 0.0290      & No  \\
         2014 AF$_{5}$   & 57000 & 1.5671674 & 0.51925929 & 6.41394   & 100.65973      & 288.73045      & 1.96                 & 0.75     & 2.38
                         & 24  & 1       & 28.80
                         & 0.4511       & 0.0450       & 0.1496      & 0.0301      & No  \\  
         2010 DU$_{1}$   & 55247 & 1.6892222 & 0.54017890 & 3.70925   & 147.83081      &  74.25332      & 2.20                 & 0.78     & 2.60
                         & 22  & 4       & 26.50 
                         & 1.0339       & 0.0187       & 0.4344      & 0.0335      & No  \\
         2013 UX         & 57000 & 1.6994971 & 0.56023315 & 5.44902   & 259.54850      &  51.72108      & 2.22                 & 0.75     & 2.65
                         & 53  & 42      & 22.00 
                         & 0.9724       & 0.0405       & 0.3799      & 0.0338      & No  \\
         2015 AM$_{45}$  & 57000 & 1.5512258 & 0.51739809 & 5.58246   & 275.22926      & 107.47502      & 1.93                 & 0.75     & 2.35
                         & 46  & 48      & 22.10 
                         & 0.4772       & 0.0338       & 0.1579      & 0.0344      & No  \\
         2014 KW$_{76}$  & 57000 & 1.6825498 & 0.56131222 & 2.33519   &  67.46190      & 102.00097      & 2.18                 & 0.74     & 2.63
                         & 34  & 4       & 27.90 
                         & 0.8000       & 0.0468       & 0.2857      & 0.0346      & No  \\
         2004 RN$_{251}$ & 57000 & 1.6558104 & 0.52789965 & 4.39200   & 179.60954      & 245.93613      & 2.13                 & 0.78     & 2.53
                         & 27  & 2       & 26.10 
                         & 0.1717       & 0.0227       & 0.0608      & 0.0360      & No  \\
         2008 EF$_{32}$  & 54536 & 1.6276459 & 0.52218191 & 1.73604   & 349.17686      & 112.26942      & 2.08                 & 0.78     & 2.48
                         & 8   & 1       & 29.40 
                         & 0.2353       & 0.0437       & 0.0767      & 0.0377      & No  \\
         2007 BD$_{7}$   & 57000 & 1.5624753 & 0.49802522 & 4.84926   & 343.62280      & 219.85002      & 1.95                 & 0.78     & 2.34
                         & 185 & 14      & 21.10 
                         & 0.9240       & 0.0440       & 0.3660      & 0.0385      & Yes \\
         2015 HR$_{43}$  & 57000 & 1.5451866 & 0.51964396 & 4.59407   & 214.18500      & 274.35526      & 1.92                 & 0.74     & 2.35
                         & 26  & 5       & 26.30 
                         & 0.4821       & 0.0246       & 0.1605      & 0.0394      & No  \\
         2011 GP$_{28}$  & 57000 & 1.5046701 & 0.50455345 & 3.81544   &  16.27364      & 256.42378      & 1.85                 & 0.75     & 2.26
                         & 14  & 1       & 29.40 
                         & 1.0278       & 0.0313       & 0.4721      & 0.0406      & No  \\
         2011 CZ$_{3}$   & 57000 & 1.5971625 & 0.51105243 & 2.11411   & 326.23428      & 241.70050      & 2.02                 & 0.78     & 2.41
                         & 30  & 4       & 26.30 
                         & 0.9529       & 0.0436       & 0.3825      & 0.0426      & No  \\
         2008 UT$_{95}$  & 57000 & 1.8149119 & 0.57453004 & 3.81156   & 220.04223      & 247.43235      & 2.45                 & 0.77     & 2.86
                         & 32  & 2       & 27.40 
                         & 0.3219       & 0.0443       & 0.1115      & 0.0449      & No  \\
         2014 OF$_{392}$ & 56867 & 1.5128759 & 0.50967476 & 4.52466   & 117.73372      & 278.59135      & 1.86                 & 0.74     & 2.28
                         & 12  & 5       & 25.50 
                         & 0.3845       & 0.0306       & 0.1286      & 0.0451      & No  \\
         2008 FH         & 54556 & 1.5849644 & 0.50476518 & 3.46186   &   5.20388      & 264.09916      & 2.00                 & 0.78     & 2.38
                         & 25  & 12      & 24.30 
                         & 1.0315       & 0.0375       & 0.4819      & 0.0453      & No  \\
         1996 AW$_{1}$   & 57000 & 1.5347586 & 0.51840437 & 4.75623   & 117.65829      & 229.26555      & 1.90                 & 0.74     & 2.33
                         & 52  & 6975    & 19.80 
                         & 0.7532       & 0.0288       & 0.2663      & 0.0458      & Yes  \\
       \hline
      \end{tabular}
      \label{candidatesCH}
      \tablenotetext{}{Note. The various $D$-criteria ($D_{\rm SH}$, $D_{\rm LS}$, $D_{\rm D}$ and $D_{\rm R}$) are also shown. The objects 
                       are sorted by ascending $D_{\rm R}$. Only objects with $D_{\rm LS} < 0.05$ and $D_{\rm R} < 0.05$ are shown. The 
                       epoch of the orbital elements is in Modified Julian Date (MJD), which is defined as the Julian date $-2400000.5$. 
                       Data as of 2015 August 19. Source: JPL Small-Body Database.
                      }
     \end{table}
     \end{landscape}
%
%
%
%
     \begin{table}
      \centering
      \fontsize{8}{11pt}\selectfont
      \tabcolsep 0.15truecm
      \caption{Heliocentric Keplerian Orbital Elements of 2003~BR$_{47}$ and 2011~EO$_{40}$ at Epoch JDCT 2457000.5 (2014 December 9.0) 
              }
      \begin{tabular}{lcc}
       \hline
                                                        &       2003~BR$_{47}$      &     2011~EO$_{40}$    \\
       \hline
        Semimajor axis, $a$ (AU)                        &   1.6283325$\pm$0.0000005 &   1.6541$\pm$0.0003   \\
        Eccentricity, $e$                               &   0.5001041$\pm$0.0000006 &   0.54016$\pm$0.00011 \\
        Inclination, $i$ (deg)                          &   4.42080$\pm$0.00002     &   3.3629$\pm$0.0007   \\
        Longitude of the ascending node, $\Omega$ (deg) & 314.56875$\pm$0.00002     &  50.303$\pm$0.008     \\
        Argument of perihelion, $\omega$ (deg)          & 112.52038$\pm$0.00012     &  17.075$\pm$0.010     \\
        Mean anomaly, $M$ (deg)                         &  274.3048$\pm$0.0009      & 312.6$\pm$0.2         \\
        Perihelion, $q$ (AU)                            &   0.8139967$\pm$0.0000008 &   0.76061$\pm$0.00006 \\
        Aphelion, $Q$ (AU)                              &   2.4426683$\pm$0.0000007 &   2.5476$\pm$0.0005   \\
        Absolute magnitude, $H$ (mag)                   &  21.4                         &  21.5             \\
       \hline
      \end{tabular}
      \label{datacandidates}
      \tablenotetext{}{Note. Values include the 1$\sigma$ uncertainty. Data as of 2015 August 19. Source: JPL Small-Body Database.
                      }
     \end{table}
%
%
%
%
%
%
%
%
%
%

\begin{figure}
        \centering
        \includegraphics[width=0.85\linewidth]{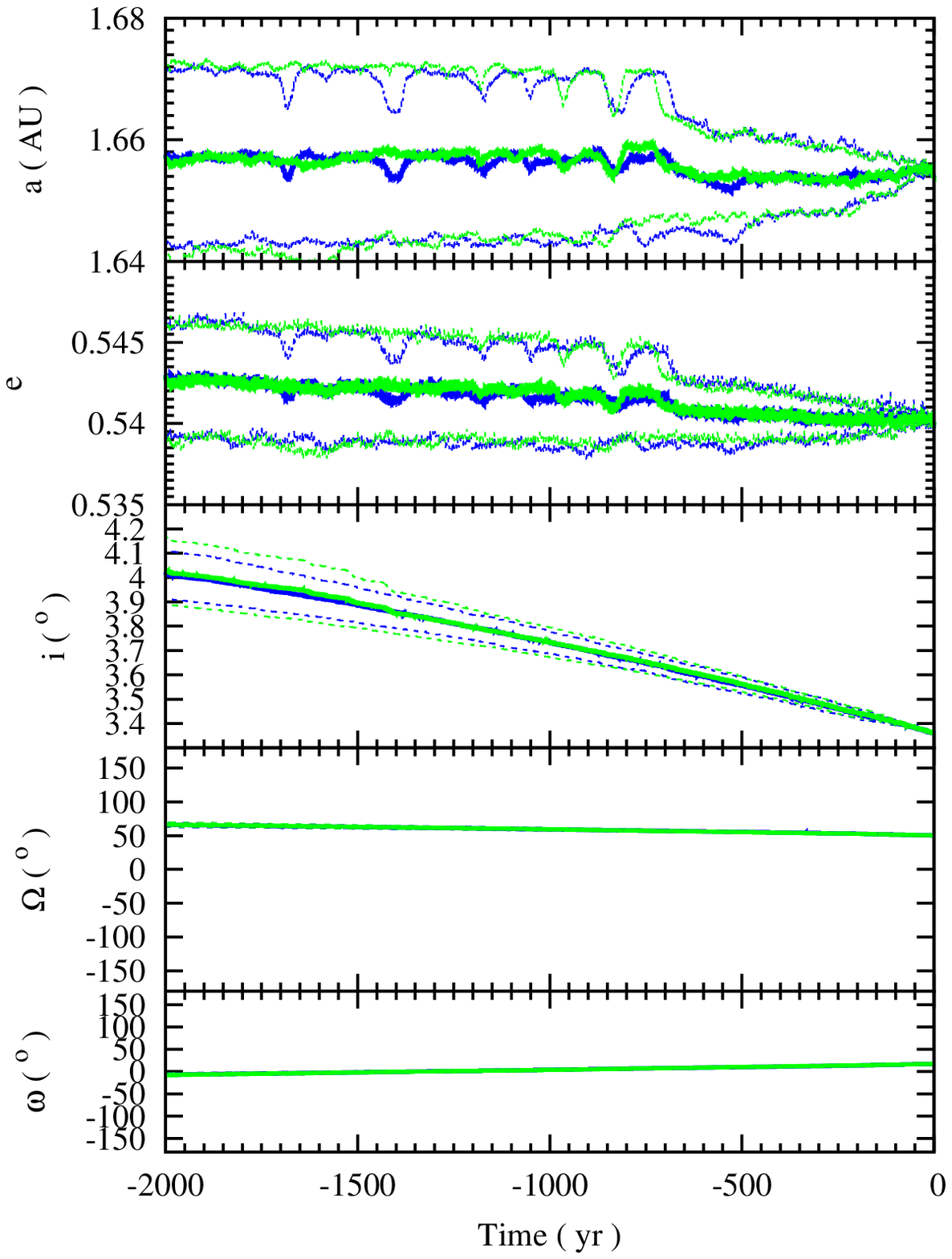}
         \caption{Time evolution of the orbital elements $a$, $e$, $i$, $\Omega$, and $\omega$ of 2011~EO$_{40}$. In blue we replot the 
                  data in Figure \ref{compare}; in green we show the results based on MCCM (see the text for details). In this figure both
                  average values and their ranges are plotted. The magnitude of the deviations is comparable to that observed when 
                  integrations are carried out with a different number of perturbing asteroids.
                 }
         \label{disper}
      \end{figure}

     As pointed out above, it may be argued that studying the orbital evolution of a given minor planet by computing orbital elements of the 
     control orbits and varying them randomly, within the ranges defined by their mean values and standard deviations, may lead to unphysical 
     results. As a consistency test, we have used the MCCM approach to recompute the past orbital evolution of 2011~EO$_{40}$, generating 
     control orbits with initial parameters from the nominal orbit, adding random noise on each initial orbital element, and making use of 
     the covariance matrix. A comparison between the results of the evolution of a sample of 100 control orbits generated using MCCM and the 
     classical method for the particular case of 2011~EO$_{40}$ appears in Figure \ref{disper}. These calculations show that, at least for 
     this particular object, the difference is not very significant; our results are therefore robust. However, and for very precise orbits, 
     the outcomes from these two approaches could be very different as we can clearly see in Figure 5 in Sitarski (1998) or in our analysis 
     of the close encounter with 367943 Duende (2012 DA$_{14}$) discussed above.
%
%
%
%

     One may also be concerned about the possible influence of the Yarkovsky and Yarkovsky--O'Keefe--Radzievskii--Paddack (YORP) effects 
     (see, e.g., Bottke et al. 2006) on our results. The largest predicted Yarkovsky drift rates are $\sim$10$^{-7}$ AU yr$^{-1}$ (see, 
     e.g., Farnocchia et al. 2013), but the gravitationally induced changes in the values of the semimajor axes of 2003~BR$_{47}$, 
     2011~EO$_{40}$, and the virtual body associated with the solution displayed in Tables \ref{orbits} and \ref{ours} are several orders of 
     magnitude larger (see Figure \ref{compare}). On the other hand, most asteroid fragments appear to be tumbling or in chaotic rotation, 
     and the role of the Yarkovsky and YORP effects may be unimportant in these cases ---but see the discussion in Vokrouhlick\'y et al. 
     (2015) for the particular case of 99942 Apophis (2004 MN$_{4}$). Besides, accurate modeling of the Yarkovsky force requires relatively 
     precise knowledge of the physical properties (for example, rotation rate, albedo, bulk density, surface conductivity, emissivity) of 
     the objects involved, which is not the case here. The non-inclusion of these effects has no major impact on the assessment completed.

     When we state that the Chelyabinsk impactor appears to be a robust dynamical relative of 2011~EO$_{40}$ we do not imply that they 
     necessarily had a physical connection in the remote past or that they have followed similar orbits in the long term even if they are
     not genetically linked. We simply state that currently, and in the immediate past, both objects appear to have been subjected to the
     same average background perturbation, i.e., that they have been sharing (in relatively recent times) the same dynamical environment. In 
     summary, they have been subjected to the same combination of secular resonances and cadence of close encounters with the objects 
     pointed out above. Schunov\'a et al. (2012) have shown that a robust statistical estimate of a dynamical relationship between objects 
     that are part of the NEA population is only possible for groups of four or more objects. On the other hand, and owing to the dynamical 
     issues described above, it is widely accepted that groups of objects moving initially in similar trajectories lose all orbital 
     coherence in a short timescale (Pauls \& Gladman 2005; Rubin \& Matson 2008; Lai et al. 2014). Focusing on the NEO population, 
     Schunov\'a et al. (2012) could not find any statistically significant group of dynamically related objects among those currently known. 
     However, Schunov\'a et al. (2014) confirmed that streams from tidally disrupted objects can be detected for a few thousand years after 
     a hypothetical disruption event only if the parent body is large enough. 

\begin{figure}
       \centering
        \includegraphics[width=\linewidth]{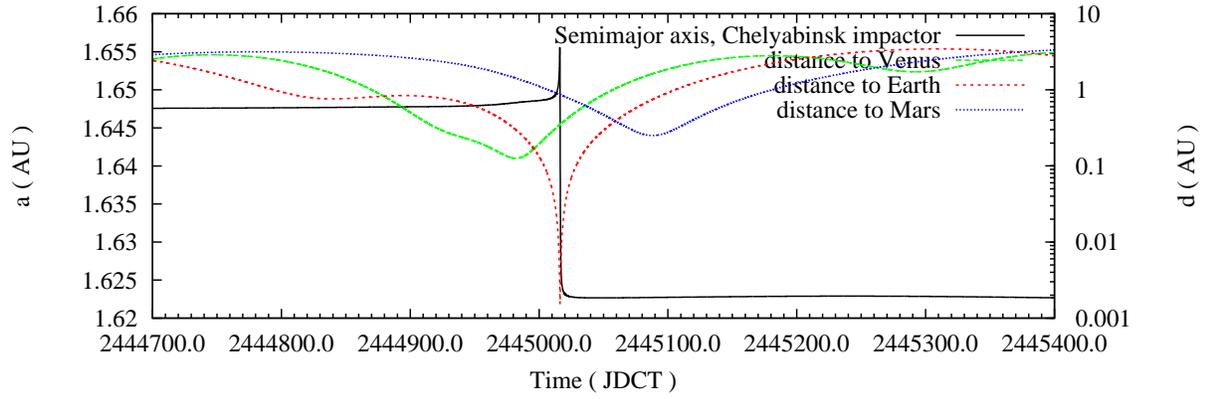}
        \caption{Time evolution of the orbital element $a$ of the Chelyabinsk impactor (black) as described by the solution displayed in 
                 Tables \ref{orbits} and \ref{ours}, and the distances to Venus (green), the Earth (red), and Mars (blue) around the time 
                 (1982 February 15) the impactor passed a gravitational keyhole that led to the impact in 2013.
                }
        \label{keyhole}
     \end{figure}

\begin{figure}
       \centering
        \includegraphics[width=\linewidth]{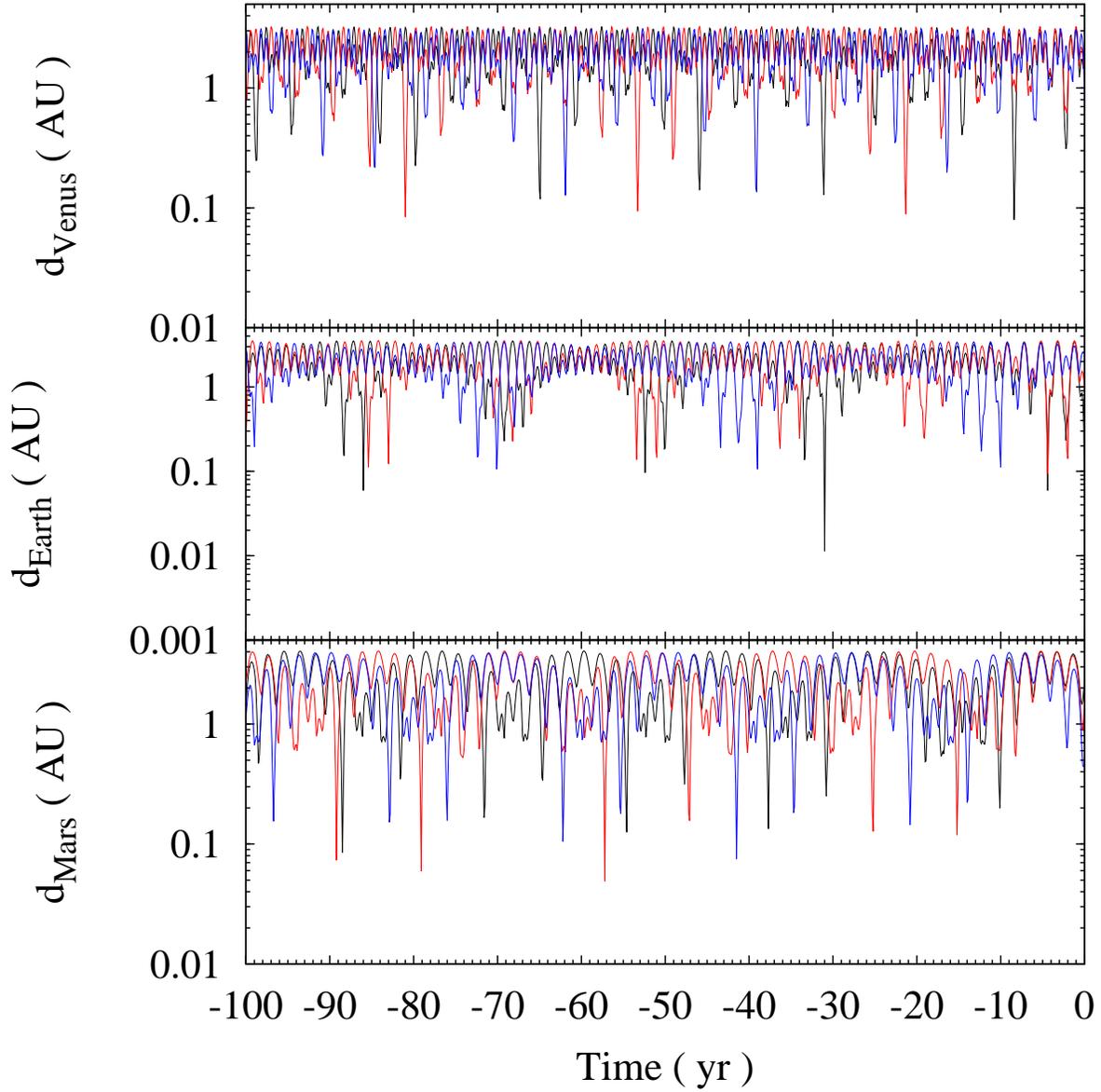}
        \caption{Distance from Venus (top panel), Earth (middle panel), and Mars (bottom panel) to the Chelyabinsk impactor (black) as 
                 described by the solution displayed in Tables \ref{orbits} and \ref{ours}, 2003~BR$_{47}$ (red), and 2011~EO$_{40}$
                 (blue). The data plotted here correspond to representative orbits from the sets displayed in Figure \ref{compare}.
                }
        \label{CE}
     \end{figure}

     Returning to the topic of the most recent events in the dynamical history of the putative Chelyabinsk impactor as characterized by 
     SOL1, all the simulations performed show that a few decades ago the object studied here suffered a dramatic change in its orbital 
     elements $a$, $e$, and $i$. Our calculations indicate that the Chelyabinsk impactor likely passed a gravitational keyhole (Chodas 1999) 
     on 1982 February 15 (2445016.294$\pm$0.018 JDCT, nominal uncertainty) during an encounter with our planet at $d <$ 0.0015 AU. As a 
     result of this close encounter, its initial 2011~EO$_{40}$-like trajectory was changed into the one that drove the meteoroid to strike 
     the Earth nearly 31 years later (see Figure \ref{keyhole}). Based solely on the number of computations performed, we estimate the 
     likelihood of this event at $>$99.9\%. As the close approaches occurred at about the same time in different years, we can speak of a 
     resonant return (Milani et al. 1999), thus the two orbits were nearly resonant as pointed out in Paper II. None of the other three 
     solutions computed (SOL0, SOL2, or SOL3) traverses a gravitational keyhole of the strength of the one found for SOL1; in particular, 
     the backwards evolution of SOL3 shows that it does not travel through any keyhole in the decades prior to its virtual impact. Only SOL1 
     produces a strong virtual resonant return. Asteroids 2011~EO$_{40}$ and 2003~BR$_{47}$ (and many others in Table \ref{candidatesCH})
     can undergo close encounters with Venus, our planet, and Mars but in general they are not synchronized or coupled in time with those of 
     the Chelyabinsk impactor (SOL1). This fact suggests that any genetic connection between the impactor and these asteroids is unlikely, 
     i.e., it cannot be a recent fragment of any of those minor bodies. However, and as we already pointed out in Paper II, 2011~EO$_{40}$
     and the Chelyabinsk impactor (SOL1) tend to encounter the Earth at somewhat regular intervals (see Figure \ref{CE}), but this could be  
     mere coincidence.
%
%
%
%
%
%
%
%
%
%
     \begin{landscape}
     \begin{table}
      \centering
      \fontsize{8}{11pt}\selectfont
      \tabcolsep 0.07truecm
      \caption{Similar to Table \ref{candidatesCH} but for SOL0. Data as of 2015 August 19. Source: JPL Small-Body Database}
      \begin{tabular}{lccccccccccccccccc}
       \hline
          Asteroid       & Epoch & $a$ (AU)  & $e$        & $i$ (deg) & $\Omega$ (deg) & $\omega$ (deg) & $P_{\rm orb}$ (year) & $q$ (AU) & $Q$ (AU) 
                         & $n$ & arc (d) & $H$ (mag) 
                         & $D_{\rm SH}$ & $D_{\rm LS}$ & $D_{\rm D}$ & $D_{\rm R}$ & PHA \\
       \hline
         2004 RU$_{109}$ & 57000 & 1.5324022 & 0.48927544 & 5.84931   & 171.40220      & 250.59089      & 1.90                 & 0.78     & 2.28
                         & 50  & 1       & 26.40
                         & 0.2020       & 0.0437       & 0.0700      & 0.0059      & No  \\
         2014 SJ$_{142}$ & 57000 & 1.5591764 & 0.50812802 & 2.08620   & 336.63875      & 119.15404      & 1.95                 & 0.77     & 2.35
                         & 34  & 30      & 23.20
                         & 0.1717       & 0.0267       & 0.0557      & 0.0061      & No  \\
         2014 DA         & 57000 & 1.5055466 & 0.48456979 & 2.46494   & 313.70691      & 110.14161      & 1.85                 & 0.78     & 2.24
                         & 142 & 21      & 22.70
                         & 0.1106       & 0.0233       & 0.0380      & 0.0145      & No  \\
         2012 VQ$_{6}$   & 56237 & 1.5074824 & 0.49522226 & 0.83915   & 211.51596      & 269.13667      & 1.85                 & 0.76     & 2.25
                         & 29  & 5       & 27.00
                         & 0.3808       & 0.0476       & 0.1257      & 0.0147      & No  \\
         2007 BD$_{7}$   & 57000 & 1.5624753 & 0.49802522 & 4.84926   & 343.62280      & 219.85002      & 1.95                 & 0.78     & 2.34
                         & 183 & 14      & 21.10
                         & 0.8928       & 0.0270       & 0.3519      & 0.0174      & Yes \\
         2011 EH         & 55622 & 1.4777343 & 0.48458290 & 2.34685   & 339.20374      &  96.57004      & 1.80                 & 0.76     & 2.19
                         & 23  & 2       & 25.50
                         & 0.0292       & 0.0264       & 0.0176      & 0.0228      & No  \\
 163679 (2002 XG$_{84}$) & 57000 & 1.4947103 & 0.47150532 & 5.05877   &  29.82729      & 349.62019      & 1.83                 & 0.79     & 2.20
                         & 144 & 4729    & 19.20
                         & 0.4720       & 0.0428       & 0.1590      & 0.0258      & No  \\
         2009 PQ$_{1}$   & 57000 & 1.4971214 & 0.48930948 & 5.59308   & 326.94431      & 271.55318      & 1.83                 & 0.76     & 2.23
                         & 81  & 29      & 22.50
                         & 0.9768       & 0.0385       & 0.4448      & 0.0267      & No  \\
         2011 CZ$_{3}$   & 57000 & 1.5971625 & 0.51105243 & 2.11411   & 326.23428      & 241.70050      & 2.02                 & 0.78     & 2.41
                         & 30  & 4       & 26.30
                         & 0.9212       & 0.0287       & 0.3692      & 0.0277      & No  \\
         2012 QZ$_{16}$  & 57000 & 1.5376728 & 0.50328326 & 6.11985   & 151.62956      & 258.93607      & 1.91                 & 0.76     & 2.31
                         & 23  & 2       & 25.50
                         & 0.2798       & 0.0467       & 0.0936      & 0.0278      & No  \\
         2008 FH         & 54556 & 1.5849644 & 0.50476518 & 3.46186   &   5.20388      & 264.09916      & 2.00                 & 0.78     & 2.38
                         & 25  & 12      & 24.30
                         & 0.9982       & 0.0148       & 0.4663      & 0.0281      & No  \\
   7753 (1988 XB)        & 57000 & 1.4676278 & 0.48187294 & 3.12289   &  73.46351      & 279.99693      & 1.78                 & 0.76     & 2.17
                         & 804 & 9214    & 18.60
                         & 0.6570       & 0.0211       & 0.2292      & 0.0283      & Yes \\
         2008 EF$_{32}$  & 54536 & 1.6276459 & 0.52218191 & 1.73604   & 349.17686      & 112.26942      & 2.08                 & 0.78     & 2.48
                         & 8   & 1       & 29.40
                         & 0.2251       & 0.0391       & 0.0756      & 0.0295      & No  \\
         2010 CJ$_{1}$   & 55239 & 1.5642628 & 0.49482798 & 4.08833   & 138.17344      & 281.78334      & 1.96                 & 0.79     & 2.34
                         & 26  & 7       & 24.10
                         & 0.1950       & 0.0219       & 0.0663      & 0.0319      & No  \\
         2015 FB$_{118}$ & 57000 & 1.4929421 & 0.46800607 & 5.12417   &   0.04734      & 106.45800      & 1.82                 & 0.79     & 2.19
                         & 46  & 18      & 23.70
                         & 0.2583       & 0.0477       & 0.0895      & 0.0322      & No  \\
         2011 EO$_{40}$  & 57000 & 1.6540887 & 0.54016233 & 3.36286   &  50.30284      &  17.07473      & 2.13                 & 0.76     & 2.55
                         & 20  & 34      & 21.50 
                         & 0.1227       & 0.0429       & 0.0549      & 0.0331      & Yes \\
         2004 RN$_{251}$ & 57000 & 1.6558104 & 0.52789965 & 4.39200   & 179.60954      & 245.93613      & 2.13                 & 0.78     & 2.53
                         & 27  & 2       & 26.10 
                         & 0.1663       & 0.0347       & 0.0634      & 0.0335      & No  \\
         2007 DL$_{41}$  & 57000 & 1.4566432 & 0.47649121 & 4.66517   & 150.81240      & 140.11206      & 1.76                 & 0.76     & 2.15
                         & 112 & 2831    & 20.70 
                         & 0.9424       & 0.0315       & 0.3958      & 0.0335      & Yes \\
         2012 TP$_{139}$ & 57000 & 1.4675833 & 0.46485745 & 2.34197   & 271.85697      &   3.26339      & 1.78                 & 0.79     & 2.15
                         & 103 & 476     & 20.30 
                         & 0.9527       & 0.0418       & 0.4328      & 0.0340      & Yes \\
         2015 CC$_{1}$   & 57070 & 1.5467360 & 0.51434097 & 3.59953   &  88.60772      & 327.84872      & 1.92                 & 0.75     & 2.34
                         & 22  & 16      & 23.88 
                         & 0.2085       & 0.0256       & 0.0704      & 0.0346      & No  \\
         2010 DU$_{1}$   & 55247 & 1.6892222 & 0.54017890 & 3.70925   & 147.83081      &  74.25332      & 2.20                 & 0.78     & 2.60
                         & 22  & 4       & 26.50 
                         & 1.0014       & 0.0419       & 0.4221      & 0.0383      & No  \\
         2013 BR$_{15}$  & 56309 & 1.5546110 & 0.52043752 & 1.95468   & 102.89879      & 284.89340      & 1.94                 & 0.75     & 2.36
                         & 10  & 2       & 25.00 
                         & 0.4305       & 0.0431       & 0.1439      & 0.0390      & No  \\
         2014 RA         & 57000 & 1.4613373 & 0.45645528 & 3.12927   & 158.00395      & 108.93260      & 1.77                 & 0.79     & 2.13
                         & 30  & 2       & 28.90
                         & 0.9593       & 0.0485       & 0.4491      & 0.0456      & No  \\
         2005 EQ$_{95}$  & 53440 & 1.6532901 & 0.52361268 & 2.37240   & 196.53230      & 251.80518      & 2.13                 & 0.79     & 2.52
                         & 17  & 2       & 23.40 
                         & 0.1447       & 0.0355       & 0.0544      & 0.0459      & No  \\
         2011 GP$_{28}$  & 57000 & 1.5046701 & 0.50455345 & 3.81544   &  16.27364      & 256.42378      & 1.85                 & 0.75     & 2.26
                         & 14  & 1       & 29.40 
                         & 0.9951       & 0.0272       & 0.4572      & 0.0463      & No  \\
         2015 AM$_{45}$  & 57000 & 1.5512258 & 0.51739809 & 5.58246   & 275.22926      & 107.47502      & 1.93                 & 0.75     & 2.35
                         & 46  & 48      & 22.10 
                         & 0.4651       & 0.0469       & 0.1549      & 0.0488      & No  \\
       \hline
      \end{tabular}
      \label{sol0}
     \end{table}
     \end{landscape}
%
%
%
%
     \begin{landscape}
     \begin{table}
      \centering
      \fontsize{8}{11pt}\selectfont
      \tabcolsep 0.07truecm
      \caption{Similar to Table \ref{candidatesCH} but for SOL2. Data as of 2015 August 19. Source: JPL Small-Body Database}
      \resizebox{\linewidth}{0.30\linewidth}{
      \begin{tabular}{lccccccccccccccccc}
       \hline
          Asteroid       & Epoch & $a$ (AU)  & $e$        & $i$ (deg) & $\Omega$ (deg) & $\omega$ (deg) & $P_{\rm orb}$ (year) & $q$ (AU) & $Q$ (AU) 
                         & $n$ & arc (d) & $H$ (mag) 
                         & $D_{\rm SH}$ & $D_{\rm LS}$ & $D_{\rm D}$ & $D_{\rm R}$ & PHA \\
       \hline
         2002 CD$_{14}$  & 57000 & 1.7779481 & 0.57944354 & 2.88679   & 127.69502      & 141.21700      & 2.37                 & 0.75     & 2.81
                         & 47  & 2549    & 20.60
                         & 1.1519       & 0.0334       & 0.5314      & 0.0088      & Yes \\
         2013 UX         & 57000 & 1.6994971 & 0.56023315 & 5.44902   & 259.54850      &  51.72108      & 2.22                 & 0.75     & 2.65
                         & 53  & 42      & 22.00 
                         & 1.0067       & 0.0171       & 0.3920      & 0.0110      & No  \\
         2006 WZ$_{3}$   & 57000 & 1.7461625 & 0.57727458 & 3.81294   & 176.32213      &   0.30728      & 2.31                 & 0.74     & 2.75
                         & 97  & 203     & 20.20
                         & 0.8975       & 0.0199       & 0.3247      & 0.0124      & No   \\
         2009 SD         & 55091 & 1.7343896 & 0.56682933 & 3.04789   & 344.32550      & 287.05721      & 2.28                 & 0.75     & 2.72
                         & 24  & 3       & 25.40 
                         & 1.1293       & 0.0304       & 0.5207      & 0.0128      & No  \\
         2012 FO$_{35}$  & 57000 & 1.8096120 & 0.57922717 & 7.15565   & 179.36210      &  93.27960      & 2.43                 & 0.76     & 2.86
                         & 27  & 15      & 23.60
                         & 1.1564       & 0.0445       & 0.5148      & 0.0134      & No  \\
         2012 VA$_{20}$  & 57000 & 1.6822360 & 0.55511183 & 4.39092   &  62.74253      & 240.11893      & 2.18                 & 0.75     & 2.62
                         & 16  & 10      & 22.80 
                         & 1.0416       & 0.0183       & 0.4194      & 0.0137      & No  \\
         2002 AC$_{9}$   & 57000 & 1.7038247 & 0.56051693 & 2.28428   &   2.58887      &  28.42304      & 2.22                 & 0.75     & 2.66
                         & 51  & 3132    & 21.00 
                         & 0.4367       & 0.0447       & 0.1432      & 0.0138      & Yes \\
         2014 KW$_{76}$  & 57000 & 1.6825498 & 0.56131222 & 2.33519   &  67.46190      & 102.00097      & 2.18                 & 0.74     & 2.63
                         & 34  & 4       & 27.90 
                         & 0.8315       & 0.0447       & 0.2958      & 0.0154      & No  \\
         2014 HQ$_{4}$   & 57000 & 1.7502026 & 0.57583344 & 6.05533   & 210.72680      & 278.76893      & 2.32                 & 0.74     & 2.76
                         & 20  & 4       & 26.20
                         & 0.5438       & 0.0237       & 0.1813      & 0.0166      & No   \\
 159677 (2002 HQ$_{11}$) & 57000 & 1.8503734 & 0.59582251 & 6.04797   & 153.33514      & 322.13054      & 2.52                 & 0.75     & 2.95
                         & 148 & 3817    & 19.40
                         & 0.4387       & 0.0327       & 0.1457      & 0.0198      & No  \\
         2015 NU$_{13}$  & 57220 & 1.8373112 & 0.58981322 & 4.20157   & 140.11078      & 259.77306      & 2.49                 & 0.75     & 2.92
                         & 21  & 10      & 19.23
                         & 0.3922       & 0.0208       & 0.1305      & 0.0200      & Yes \\
         2013 OW$_{2}$   & 57000 & 1.8307041 & 0.59311032 & 6.50010   & 221.75879      & 336.76736      & 2.48                 & 0.74     & 2.92
                         & 22  & 28      & 21.10
                         & 1.0357       & 0.0371       & 0.4010      & 0.0215      & No   \\
 358471 (2007 NS$_{4}$)  & 57000 & 1.8747202 & 0.59757148 & 5.80175   &  11.23176      &  47.25005      & 2.57                 & 0.75     & 2.99
                         & 83  & 2061    & 19.10
                         & 0.1945       & 0.0319       & 0.0654      & 0.0227      & Yes \\
         2001 WM$_{15}$  & 52239 & 1.8653862 & 0.59394567 & 5.41106   & 244.20729      & 250.90778      & 2.55                 & 0.76     & 2.97
                         & 34  & 4       & 25.00
                         & 0.5903       & 0.0262       & 0.1982      & 0.0241      & No  \\
         2005 TE         & 53646 & 1.7473215 & 0.57690200 & 6.48704   &  13.07234      & 270.57774      & 2.31                 & 0.74     & 2.76
                         & 17  & 6       & 23.90 
                         & 1.1191       & 0.0318       & 0.4871      & 0.0243      & No  \\
         1996 VB$_{3}$   & 50401 & 1.6259062 & 0.54444404 & 2.79583   & 180.62271      & 132.68222      & 2.07                 & 0.74     & 2.51
                         & 21  & 9       & 22.40 
                         & 0.9877       & 0.0447       & 0.3823      & 0.0265      & No  \\
  86039 (1999 NC$_{43}$) & 57000 & 1.7595889 & 0.57919044 & 7.12356   & 311.81453      & 120.57423      & 2.33                 & 0.74     & 2.78
                         & 891 & 5492    & 16.00
                         & 0.0606       & 0.0426       & 0.0211      & 0.0269      & Yes \\
         2011 GZ$_{2}$   & 57000 & 1.8367961 & 0.58046296 & 6.99505   &  18.57475      & 247.21411      & 2.49                 & 0.77     & 2.90
                         & 51  & 6       & 26.10 
                         & 1.1529       & 0.0459       & 0.5460      & 0.0290      & No  \\
 267729 (2003 FC$_{5}$)  & 57000 & 1.9167576 & 0.60910814 & 5.82513   & 189.25854      & 270.65349      & 2.65                 & 0.75     & 3.08
                         & 629 & 4716    & 18.30
                         & 0.3035       & 0.0414       & 0.1062      & 0.0291      & Yes \\
 199801 (2007 AE$_{12}$) & 57000 & 1.6843105 & 0.56995521 & 2.28470   & 245.70900      &  86.67112      & 2.19                 & 0.72     & 2.64
                         & 138 & 4995    & 19.50
                         & 0.9009       & 0.0493       & 0.3294      & 0.0326      & Yes \\
         2014 TM$_{35}$  & 56948 & 1.9283432 & 0.60990027 & 3.20973   & 215.35752      &  87.14549      & 2.68                 & 0.75     & 3.10
                         & 40  & 13      & 23.70
                         & 1.0920       & 0.0465       & 0.4390      & 0.0333      & No  \\
         2008 EW$_{84}$  & 54538 & 1.7504903 & 0.58321185 & 5.34640   & 170.56633      &  82.95460      & 2.32                 & 0.73     & 2.77
                         & 16  & 2       & 24.50
                         & 1.1682       & 0.0239       & 0.5483      & 0.0333      & No  \\
         2011 EO$_{40}$  & 57000 & 1.6540887 & 0.54016233 & 3.36286   &  50.30284      &  17.07473      & 2.13                 & 0.76     & 2.55
                         & 20  & 34      & 21.50
                         & 0.1323       & 0.0421       & 0.0516      & 0.0336      & Yes \\
         2007 EZ         & 57000 & 1.7070245 & 0.57260188 & 5.81095   &  98.52168      & 357.87061      & 2.23                 & 0.73     & 2.68
                         & 111 & 3078    & 19.80 
                         & 0.2636       & 0.0261       & 0.0865      & 0.0365      & No  \\
         2010 TV$_{54}$  & 57000 & 1.9133256 & 0.61405795 & 6.18704   & 202.37865      & 254.99077      & 2.65                 & 0.74     & 3.09
                         & 31  & 6       & 25.70
                         & 0.2831       & 0.0497       & 0.1015      & 0.0374      & No  \\
         2012 BC$_{77}$  & 57000 & 1.8889272 & 0.59441053 & 4.60747   & 149.04739      &  58.14416      & 2.60                 & 0.77     & 3.01
                         & 26  & 9       & 25.10
                         & 1.0757       & 0.0287       & 0.4266      & 0.0402      & No  \\
         2011 GC$_{55}$  & 57000 & 1.8242312 & 0.58019970 & 2.77258   &  98.96323      &  22.41153      & 2.46                 & 0.77     & 2.88
                         & 62  & 24      & 23.00
                         & 0.4612       & 0.0397       & 0.1512      & 0.0413      & No  \\
         2012 PS$_{4}$   & 57000 & 1.7676567 & 0.59240205 & 3.93689   &  86.25805      & 144.14991      & 2.35                 & 0.72     & 2.81
                         & 67  & 36      & 21.20 
                         & 1.1437       & 0.0370       & 0.4995      & 0.0419      & Yes \\
         2010 TW$_{149}$ & 55484 & 1.8922324 & 0.61598905 & 4.57001   & 195.85511      & 103.38629      & 2.60                 & 0.73     & 3.06
                         & 20  & 1       & 25.90
                         & 1.1146       & 0.0488       & 0.4524      & 0.0434      & No  \\
 220839 (2004 VA)        & 57000 & 1.9004546 & 0.59634469 & 3.69800   & 109.66945      &  43.10495      & 2.62                 & 0.77     & 3.03
                         & 333 & 4076    & 17.30
                         & 0.7398       & 0.0358       & 0.2538      & 0.0462      & Yes \\
         2008 UT$_{95}$  & 57000 & 1.8149119 & 0.57453004 & 3.81156   & 220.04223      & 247.43235      & 2.45                 & 0.77     & 2.86
                         & 32  & 2       & 27.40
                         & 0.3366       & 0.0293       & 0.1114      & 0.0480      & No  \\
       \hline
      \end{tabular}
      }
      \label{sol2}
     \end{table}
     \end{landscape}
%
%
%
%
     \begin{landscape}
     \begin{table}
      \centering
      \fontsize{8}{11pt}\selectfont
      \tabcolsep 0.07truecm
      \caption{Similar to Table \ref{candidatesCH} but for SOL3. Data as of 2015 August 19. Source: JPL Small-Body Database}
      \resizebox{\linewidth}{0.30\linewidth}{
      \begin{tabular}{lccccccccccccccccc}
       \hline
          Asteroid       & Epoch & $a$ (AU)  & $e$        & $i$ (deg) & $\Omega$ (deg) & $\omega$ (deg) & $P_{\rm orb}$ (year) & $q$ (AU) & $Q$ (AU) 
                         & $n$ & arc (d) & $H$ (mag) 
                         & $D_{\rm SH}$ & $D_{\rm LS}$ & $D_{\rm D}$ & $D_{\rm R}$ & PHA \\
       \hline
         2013 OW$_{2}$   & 57000 & 1.8307041 & 0.59311032 & 6.50010   & 221.75879      & 336.76736      & 2.48                 & 0.74     & 2.92
                         & 22  & 28      & 21.10
                         & 1.0592       & 0.0225       & 0.4099      & 0.0011      & No   \\
 159677 (2002 HQ$_{11}$) & 57000 & 1.8503734 & 0.59582251 & 6.04797   & 153.33514      & 322.13054      & 2.52                 & 0.75     & 2.95
                         & 148 & 3817    & 19.40
                         & 0.4514       & 0.0158       & 0.1490      & 0.0084      & No  \\
         2010 TV$_{54}$  & 57000 & 1.9133256 & 0.61405795 & 6.18704   & 202.37865      & 254.99077      & 2.65                 & 0.74     & 3.09
                         & 31  & 6       & 25.70
                         & 0.2906       & 0.0236       & 0.0997      & 0.0151      & No  \\
         2014 HQ$_{4}$   & 57000 & 1.7502026 & 0.57583344 & 6.05533   & 210.72680      & 278.76893      & 2.32                 & 0.74     & 2.76
                         & 20  & 4       & 26.20
                         & 0.5591       & 0.0254       & 0.1872      & 0.0156      & No   \\
         2006 WZ$_{3}$   & 57000 & 1.7461625 & 0.57727458 & 3.81294   & 176.32213      &   0.30728      & 2.31                 & 0.74     & 2.75
                         & 97  & 203     & 20.20
                         & 0.9194       & 0.0321       & 0.3329      & 0.0168      & No   \\
 267729 (2003 FC$_{5}$)  & 57000 & 1.9167576 & 0.60910814 & 5.82513   & 189.25854      & 270.65349      & 2.65                 & 0.75     & 3.08
                         & 629 & 4716    & 18.30
                         & 0.3124       & 0.0181       & 0.1061      & 0.0172      & Yes \\
  86039 (1999 NC$_{43}$) & 57000 & 1.7595889 & 0.57919044 & 7.12356   & 311.81453      & 120.57423      & 2.33                 & 0.74     & 2.78
                         & 891 & 5492    & 16.00
                         & 0.0565       & 0.0371       & 0.0231      & 0.0175      & Yes \\
         2005 TE         & 53646 & 1.7473215 & 0.57690200 & 6.48704   &  13.07234      & 270.57774      & 2.31                 & 0.74     & 2.76
                         & 17  & 6       & 23.90 
                         & 1.1428       & 0.0294       & 0.4973      & 0.0181      & No  \\
 358471 (2007 NS$_{4}$)  & 57000 & 1.8747202 & 0.59757148 & 5.80175   &  11.23176      &  47.25005      & 2.57                 & 0.75     & 2.99
                         & 83  & 2061    & 19.10
                         & 0.1956       & 0.0171       & 0.0630      & 0.0214      & Yes \\
         2010 TW$_{149}$ & 55484 & 1.8922324 & 0.61598905 & 4.57001   & 195.85511      & 103.38629      & 2.60                 & 0.73     & 3.06
                         & 20  & 1       & 25.90
                         & 1.1367       & 0.0263       & 0.4596      & 0.0215      & No  \\
         2008 EW$_{84}$  & 54538 & 1.7504903 & 0.58321185 & 5.34640   & 170.56633      &  82.95460      & 2.32                 & 0.73     & 2.77
                         & 16  & 2       & 24.50
                         & 1.1939       & 0.0175       & 0.5588      & 0.0231      & No  \\
         2012 FO$_{35}$  & 56013 & 1.8096120 & 0.57922717 & 7.15565   & 179.36210      &  93.27960      & 2.43                 & 0.76     & 2.86
                         & 27  & 15      & 23.60
                         & 1.1818       & 0.0432       & 0.5248      & 0.0236      & No  \\
         2002 CD$_{14}$  & 57000 & 1.7779481 & 0.57944354 & 2.88679   & 127.69502      & 141.21700      & 2.37                 & 0.75     & 2.81
                         & 47  & 2549    & 20.60
                         & 1.1772       & 0.0455       & 0.5419      & 0.0250      & Yes \\
         2013 VX$_{13}$  & 56610 & 1.9595239 & 0.62550101 & 6.14648   &  17.37851      & 301.55846      & 2.74                 & 0.73     & 3.19
                         & 24  & 14      & 22.10
                         & 1.0461       & 0.0331       & 0.3988      & 0.0252      & No  \\
         2009 MG$_{1}$   & 55007 & 2.0038233 & 0.62552004 & 6.79624   &  87.89116      & 101.43854      & 2.84                 & 0.75     & 3.26
                         & 19  & 11      & 24.00 
                         & 1.0385       & 0.0405       & 0.3882      & 0.0262      & No  \\
         2012 SV$_{9}$   & 56190 & 2.0097244 & 0.63303198 & 4.79202   & 188.73213      &  80.70820      & 2.85                 & 0.74     & 3.28
                         & 18  & 5       & 23.90 
                         & 1.2324       & 0.0370       & 0.5610      & 0.0263      & No  \\
         2015 NU$_{13}$  & 57220 & 1.8373112 & 0.58981322 & 4.20157   & 140.11078      & 259.77306      & 2.49                 & 0.75     & 2.92
                         & 21  & 10      & 19.23
                         & 0.4006       & 0.0239       & 0.1332      & 0.0277      & Yes \\
         2013 UX         & 57000 & 1.6994971 & 0.56023315 & 5.44902   & 259.54850      &  51.72108      & 2.22                 & 0.75     & 2.65
                         & 53  & 42      & 22.00 
                         & 1.0285       & 0.0376       & 0.4009      & 0.0280      & No  \\
         2001 WM$_{15}$  & 52239 & 1.8653862 & 0.59394567 & 5.41106   & 244.20729      & 250.90778      & 2.55                 & 0.76     & 2.97
                         & 34  & 4       & 25.00
                         & 0.6054       & 0.0177       & 0.2028      & 0.0284      & No  \\
         2012 PS$_{4}$   & 57000 & 1.7676567 & 0.59240205 & 3.93689   &  86.25805      & 144.14991      & 2.35                 & 0.72     & 2.81
                         & 67  & 36      & 21.20 
                         & 1.1687       & 0.0307       & 0.5100      & 0.0284      & Yes \\
         2012 HD$_{25}$  & 56046 & 1.9765398 & 0.63026934 & 5.95173   &  42.63974      &  93.21000      & 2.78                 & 0.73     & 3.22
                         & 47  & 5       & 23.60
                         & 0.6255       & 0.0367       & 0.2094      & 0.0296      & No  \\
 303449 (2005 BE$_{2}$)  & 57000 & 2.0152022 & 0.62594237 & 6.57681   &  92.63080      & 162.40881      & 2.86                 & 0.75     & 3.28
                         & 276 & 3600    & 18.20
                         & 1.2371       & 0.0394       & 0.5912      & 0.0303      & No  \\
         2007 EZ         & 57000 & 1.7070245 & 0.57260188 & 5.81095   &  98.52168      & 357.87061      & 2.23                 & 0.73     & 2.68
                         & 111 & 3078    & 19.80 
                         & 0.2742       & 0.0283       & 0.0917      & 0.0309      & No  \\
         2014 TM$_{35}$  & 56948 & 1.9283432 & 0.60990027 & 3.20973   & 215.35752      &  87.14549      & 2.68                 & 0.75     & 3.10
                         & 40  & 13      & 23.70
                         & 1.1139       & 0.0398       & 0.4464      & 0.0323      & No  \\
         2012 VA$_{20}$  & 57000 & 1.6822360 & 0.55511183 & 4.39092   &  62.74253      & 240.11893      & 2.18                 & 0.75     & 2.62
                         & 16  & 10      & 22.80 
                         & 1.0649       & 0.0453       & 0.4294      & 0.0346      & No  \\
         2002 VR$_{85}$  & 57000 & 1.8169080 & 0.60420879 & 6.01821   & 204.85943      & 298.80545      & 2.45                 & 0.72     & 2.91
                         & 59  & 29      & 20.40
                         & 0.6955       & 0.0260       & 0.2363      & 0.0382      & Yes \\
         2011 GZ$_{2}$   & 57000 & 1.8367961 & 0.58046296 & 6.99505   &  18.57475      & 247.21411      & 2.49                 & 0.77     & 2.90
                         & 51  & 6       & 26.10 
                         & 1.1777       & 0.0460       & 0.5576      & 0.0399      & No  \\
    4183 Cuno (1959 LM)  & 57000 & 1.9822062 & 0.63441443 & 6.70752   & 294.93347      & 236.27353      & 2.79                 & 0.72     & 3.24
                         & 1645& 20497   & 14.40
                         & 0.9159       & 0.0478       & 0.3296      & 0.0414      & Yes \\
         2015 PK         & 57000 & 1.9296505 & 0.59917954 & 7.21027   & 133.28762      & 111.31405      & 2.68                 & 0.77     & 3.09
                         & 26  & 1       & 28.14 
                         & 1.2104       & 0.0476       & 0.5496      & 0.0458      & No  \\
         2013 EB$_{34}$  & 56364 & 1.8594018 & 0.61304902 & 7.33262   & 160.04942      &  96.99762      & 2.54                 & 0.72     & 3.00
                         & 20  & 4       & 24.70
                         & 1.2297       & 0.0447       & 0.5682      & 0.0460      & No  \\
         2013 EF$_{68}$  & 56365 & 1.9812035 & 0.63801017 & 5.56067   & 177.16206      & 277.12496      & 2.79                 & 0.72     & 3.25
                         & 28  & 1       & 25.80
                         & 0.2751       & 0.0474       & 0.1003      & 0.0462      & No  \\
         2013 VW$_{13}$  & 57000 & 1.6726846 & 0.57435835 & 3.52659   & 227.74983      & 101.32453      & 2.16                 & 0.71     & 2.63
                         & 34  & 21      & 26.20
                         & 0.9463       & 0.0472       & 0.3488      & 0.0480      & No  \\
         2012 BC$_{77}$  & 57000 & 1.8889272 & 0.59441053 & 4.60747   & 149.04739      &  58.14416      & 2.60                 & 0.77     & 3.01
                         & 26  & 9       & 25.10
                         & 1.1002       & 0.0284       & 0.4360      & 0.0482      & No  \\
         2014 TK$_{64}$  & 56948 & 1.9634840 & 0.60692881 & 6.72030   &  16.22175      &  88.43995      & 2.75                 & 0.77     & 3.16
                         & 17  & 7       & 23.30
                         & 0.3141       & 0.0418       & 0.1030      & 0.0484      & No  \\
       \hline
      \end{tabular}
      }
      \label{sol3}
     \end{table}
     \end{landscape}
%
%

  \section{Discussion and conclusions}
     In this paper, we have obtained a statistically robust solution for the pre-impact orbit of the Chelyabinsk impactor. This solution has 
     been computed by making use of full $N$-body calculations and it reproduces, within very narrow limits ($<$0.0001\% in time and 
     $<$0.4\% in position), the well documented values of both impact time and location of the Chelyabinsk event. It is compatible with 
     other observational properties as well. The impact probability of our solution is $>$~99.999\% and it is consistent with those in 
     Papers I and II although the methodology behind it is completely different. It also matches well the one originally computed by S. 
     Nakano:$^5$ relative differences of 0.14\% in $a$, 0.14\% in $e$, 2.7\% in $i$, 0.011\% in $\Omega$, and 0.007\% in $\omega$. Our 
     simulations also confirm the existence of a reasonably strong dynamical link between the PHA 2011~EO$_{40}$ and the Chelyabinsk 
     impactor as described by the solution in Table \ref{ours}. Alternative, relevant candidate solutions are also explored. Our statistical 
     analysis shows that the value of the geocentric velocity of the impactor at the entry point in the atmosphere is currently the key 
     limiting parameter to obtaining a robust and final determination of the orbital solution of the Chelyabinsk asteroid. Further work may 
     be required in that respect. On the other hand, our study vindicates the role of quality control based on $N$-body-integrations in the 
     determination of pre-impact meteor orbits. Without such quality control, orbital solutions may be meaningless as they do not produce 
     any relevant impacts. A statistical analysis should be standard practice in these cases.

     Borovi\v{c}ka et al. (2013) suggested that the Chelyabinsk impactor and the PHA 86039 (1999~NC$_{43}$) were once part of the same 
     object. Reddy et al. (2015) later pointed out that the existence of a connection between the Chelyabinsk meteoroid and 86039 is very 
     weak, both in dynamical and compositional terms. Here, we lend further support to this conclusion. Our extensive simulations show that 
     it is highly unlikely that, prior to colliding with our planet, the Chelyabinsk impactor followed an orbit similar to the ones 
     described in Borovi\v{c}ka et al. (2013) or Popova et al. (2013). In both cases, their orbital solutions are unable to place the 
     impactor sufficiently close to our planet within 10 or less minutes of the well documented value of the impact time, which is 
     consistent with our analysis in Paper II.

     On 1908 June 30 a small body was observed streaking across the daytime sky in a remote part of Russia, above the Tunguska River. The
     subsequent meteor airburst, known as the Tunguska event, is considered the most powerful asteroid impact witnessed to date (see, e.g.,
     Farinella et al. 2001). Almost 107 years later, the Chelyabinsk event has become the second most powerful instance of an observed 
     meteor airburst (Brown et al. 2013; Le Pichon et al. 2013). In both cases, the glare of the Sun provided an effective hiding spot to 
     the eventual impactor, highlighting the fact that these objects are still a challenge for our modern resources. However, asteroids 
     impacting from the direction of the Sun are probably no different from those impacting at opposition (Hills \& Leonard 1995) although 
     they could form more than 30\% of the NEOs (Isobe \& Yoshikawa 1997) and are better detected from spaceborne telescopes, perhaps 
     located at the L$_2$ point or orbiting Venus (Hills 1992; Hills \& Leonard 1995). A number of initiatives in that direction have been 
     discussed recently (e.g., Dunham et al. 2013; Mainzer et al. 2015). 
 
  \acknowledgments
     We thank the anonymous referee for a constructive and helpful report, E. Schunov\'a for her feedback on earlier versions of this 
     manuscript, S. D. Miller for comments on his results, and P. Wiegert for sharing the details of his pre-impact orbital solution. This 
     work was partially supported by the Spanish ``Comunidad de Madrid'' under grant CAM S2009/ESP-1496. Some of the calculations discussed 
     here were completed on the ``Servidor Central de C\'alculo'' of the Universidad Complutense de Madrid. In the preparation of this 
     paper, we made use of the NASA Astrophysics Data System, the ASTRO-PH e-print server, the MPC data server, and the NEODyS information 
     service.

  \newpage
  \appendix
  \section{Cartesian state vectors at epoch JDCT 2456337.638888889 = A.D. 2013 February 14 03:20:00.0000 UTC}
     In order to facilitate verification of our results by other astrodynamicists, we show in Table \ref{Cartesian} the Cartesian state 
     vectors of the physical model used in all the calculations presented here. These values have been computed by the SSDG, Horizons 
     On-Line Ephemeris System at epoch JDCT 2456337.638888889 = A.D. 2013 February 14 03:20:00.0000 UTC; this instant is considered as 
     $t = 0$ across this work unless explicitly stated. Positions and velocities are referred to the barycenter of the solar system. 
%
%
     \begin{landscape}
     \begin{table}
      \centering
      \fontsize{8}{11pt}\selectfont
      \tabcolsep 0.06truecm
      \caption{Cartesian State Vectors at Epoch JDCT 2456337.638888889 = A.D. 2013 February 14 03:20:00.0000 UTC (Source: JPL \textsc{Horizons} 
               System, Data as of 2015 April 27)}
      \resizebox{\linewidth}{0.30\linewidth}{
      \begin{tabular}{lccccccc}
       \hline
          Body              & Mass (kg)    &
          $X$ (AU)          & $Y$ (AU)            & $Z$ (AU)             & $V_{\rm X}$ (AU/day)  & $V_{\rm Y}$ (AU/day)  & $V_{\rm Z}$ (AU/day)       \\
       \hline
          Sun                     & 1.988544D+30 &
          -1.011494946041879D-03 & -2.504114162880022D-03 & -4.792865115835242D-05 &  6.220894815132611D-06 & -8.990667125614989D-07 & -1.353378656456761D-07 \\
          Mercury                 & 3.302D+23    &
           1.601238578327888D-01 &  2.625722472443697D-01 &  6.825890471667526D-03 & -2.964378655400363D-02 &  1.573883688706671D-02 &  4.006348641371026D-03 \\
          Venus                   & 48.685D+23   &
           3.684190336612509D-01 & -6.292530937012436D-01 & -2.995683700834373D-02 &  1.729515787108521D-02 &  1.019957367549420D-02 & -8.581635029629278D-04 \\
          Earth                   & 5.97219D+24  &
          -8.140698930244055D-01 &  5.581120070644801D-01 & -6.589763331176098D-05 & -1.003900706480297D-02 & -1.423531287380917D-02 &  7.820902130810397D-07 \\
          Moon                    & 734.9D+20    & 
          -8.115470135393000D-01 &  5.587246391320240D-01 &  7.135598656712641D-05 & -1.014006812230196D-02 & -1.366171847421980D-02 & -3.819647952433072D-05 \\
          Mars                    & 6.4185D+23   & 
           1.358696919954038D+00 & -2.609405990577065D-01 & -3.884700972410955D-02 &  3.155433071794785D-03 &  1.494047053495886D-02 &  2.355938303052990D-04 \\
          (1) Ceres               & 8.958D+20    &
          -4.530641758590559D-01 &  2.582498614052824D+00 &  1.645168996062354D-01 & -1.039618361109546D-02 & -2.573247707241392D-03 &  1.837093277694028D-03 \\
          (2) Pallas              & 2.108D+20    &
           2.148722377773673D+00 &  1.139324612315395D+00 & -9.682180658594401D-01 & -8.045289972243500D-03 &  6.359998630354690D-03 & -3.724517164654106D-03 \\
          (4) Vesta               & 2.59076D+20  &
          -1.208190498587545D-01 &  2.553051597561101D+00 & -6.210372692193103D-02 & -1.018543503227075D-02 & -7.531638828710834D-04 &  1.261808261070763D-03 \\
          (10) Hygiea             & 8.67D+19     &
           3.292138218259053D+00 & -7.753641456595389D-02 &  2.138959793656113D-01 &  1.160053746471100D-03 &  9.141540670241449D-03 &  2.177016629356885D-04 \\
          (31) Euphrosyne         & 5.81D+19     &
          -2.602099194748581D+00 &  5.537190212892845D-01 &  9.009030329104319D-01 & -5.116957924612657D-03 & -9.168433478330714D-03 & -2.569751792414020D-03 \\
          Jupiter                 & 1898.13D+24  & 
           1.097408567386605D+00 &  4.954902903200203D+00 & -4.521676703255616D-02 & -7.458634311741788D-03 &  1.992263681815774D-03 &  1.586575650167676D-04 \\
          Saturn                  & 5.68319D+26  &
          -7.954280583549062D+00 & -5.719387687644644D+00 &  4.160061774451558D-01 &  2.953857543128144D-03 & -4.543457016907906D-03 & -3.820665181298168D-05 \\
          Uranus                  & 86.8103D+24  &
           1.986481516054372D+01 &  2.734885430030593D+00 & -2.472007163305052D-01 & -5.653198932594880D-04 &  3.713021339159490D-03 &  2.101603062461898D-05 \\
          Neptune                 & 102.41D+24   &
           2.662268720316427D+01 & -1.380194387354347D+01 & -3.293215144108281D-01 &  1.424320769695806D-03 &  2.805857995688192D-03 & -9.052860825176624D-05 \\
          Pluto-Charon barycenter & 1.45712D+22  &
           5.247177290509663D+00 & -3.190345013616387D+01 &  1.896073175017724D+00 &  3.157371987049258D-03 & -1.182716627674995D-04 & -9.006421307059045D-04 \\
       \hline
          Nominal impactor        & 1.0D+07      &
          -8.0679626173D-01      &  5.5488529365D-01      &  1.29912462D-03        & -1.726521543D-02       & -1.104192256D-02       & -1.30271852D-03        \\
       \hline
      \end{tabular}
      }
      \label{Cartesian}
      \tablenotetext{}{Note. The sample Cartesian vector for the Chelyabinsk impactor corresponds to the nominal orbit in Table \ref{ours}.
                      }
     \end{table}
     \end{landscape}
%
%

\end{document}